\documentclass[acmsmall,screen,nonacm]{acmart}
\usepackage{booktabs}   %
\usepackage{subcaption} %

\usepackage[export]{adjustbox}
\usepackage[normalem]{ulem}
\usepackage[scaled=0.8]{beramono}
\usepackage{placeins}
\usepackage[skip=2.5pt]{caption}
\usepackage{bcprules}
\usepackage{bm}
\usepackage{cancel}
\usepackage{diagbox}
\usepackage{enumitem}
\usepackage{etoolbox}
\usepackage{float}
\usepackage{hyperref}
\usepackage{lipsum}
\usepackage{listings,array,varwidth}
\usepackage{makecell}
\usepackage{mathpartir}
\usepackage{amsmath}
\usepackage{mathtools}
\usepackage{mdframed}
\usepackage{multirow}
\usepackage{multifootnote}
\usepackage{scalerel}
\usepackage{soul}
\usepackage{tabularx}
\usepackage{tikz-cd}
\usepackage{tikz} %
\usepackage{varioref}
\usepackage{wrapfig}
\usepackage{xcolor,colortbl}
\usepackage{xfrac}
\usepackage{xifthen}
\usepackage{xspace}
\usepackage{xurl}
\usetikzlibrary{arrows, arrows.meta, shapes, positioning}
\tikzset{nodearrow/.style={black, ->, >=myarrow},
myarrow/.tip={Latex[width=1mm, length=1mm]},
tracked/.style={draw=black, fill=blue!10},
locs/.style={draw, circle, tracked},
shared/.style={fill=yellow!20},
circular locs/.style={locs, fill=purple!10},
shared locs/.style={locs, shared},
lambda/.style={draw, cloud, text centered, cloud puffs=15, aspect=2.5},
untracked lambda/.style={lambda, fill=gray!10, dash pattern=on 5pt off 2pt},
tracked lambda/.style={lambda, tracked},
shared lambda/.style={lambda, shared},
}

\usepackage{cleveref}
\usepackage{pifont} %

\setlist[enumerate, 1]{%
  leftmargin = 1.2\parindent, %
  align = left,
  labelwidth=\parindent,
  labelsep = 1pt
}
\lstdefinelanguage{DOT}%
{morekeywords={val,new},%
  sensitive,%
  morecomment=[l]//,%
  morecomment=[s]{/*}{*/},%
  morestring=[b]",%
  morestring=[b]',%
  showstringspaces=false%
}[keywords,comments,strings]%

\newlength{\trulemargin}
\newlength{\trulewidth}
\newlength{\srulewidth}
\newenvironment{trules}{$\vspace{0.5em}\ba{p{\trulemargin}@{~}p{\trulewidth}@{~}p{\trulemargin}}}{\ea$}
\newenvironment{srules}{$\vspace{0.5em}\ba{p{\trulemargin}@{~}p{\srulewidth}}}{\ea$}

\newcommand{\ba}{\begin{array}}
\newcommand{\ea}{\end{array}}

\newcommand{\ei}{\end{array}}
\newcommand{\bcases}{\left\{\begin{array}{ll}}
\newcommand{\ecases}{\end{array}\right.}

\newcommand{\eg}{{\em e.g.}\xspace}
\newcommand{\ie}{{\em i.e.}\xspace}

\newcommand{\dom}{\mbox{\sl dom}}

\newcommand{\judgement}[2]{{\textsf{\textbf{#1}}} \hfill #2}

\usepackage{listings,array,varwidth}
\usepackage{mathpartir}
\usepackage{mathtools}
\usepackage{makecell}

\usepackage{diagbox}
\usepackage{float}
\usepackage{varioref}
\usepackage{xfrac}
\usepackage{enumitem}
\usepackage[skip=2.5pt]{caption}

\usepackage{xifthen}

\usepackage{tikz}
\usepackage{xsavebox}
\usetikzlibrary{arrows}

\makeatletter %
\def\arcr{\@arraycr}
\makeatother

\newcommand{\showDOI}[1]{\unskip}

\newtheorem{innercustomgeneric}{\customgenericname}
\providecommand{\customgenericname}{}
\newcommand{\newcustomtheorem}[2]{%
  \newenvironment{#1}[1]
  {%
   \renewcommand\customgenericname{#2}%
   \renewcommand\theinnercustomgeneric{##1}%
   \innercustomgeneric
  }
  {\endinnercustomgeneric}
}
\newcustomtheorem{re-definition}{Definition}
\newcustomtheorem{re-lemma}{Lemma}

\makeatletter
\DeclareRobustCommand{\etc}{%
    \@ifnextchar{.}%
        {etc}%
        {etc.\@\xspace}%
}
\makeatother

\newcommand{\rch}{\textit{one-step store reachable}}
\newcommand{\rchty}{\textit{one-step store reachability}}

\newcommand{\Specsharp}{%
	{\settoheight{\dimen0}{C}Spec\kern-.05em \resizebox{!}{\dimen0}{\raisebox{\depth}{\#}}}}
\newcommand{\Csharp}{%
	{\settoheight{\dimen0}{C}C\kern-.05em \resizebox{!}{\dimen0}{\raisebox{\depth}{\#}}}}

\newcommand{\fun}[1]{\operatorname{#1}}
\newcommand{\DOM}{\fun{dom}}

\definecolor{blue-violet}{rgb}{0.54, 0.17, 0.89}
\definecolor{dark-cyan}{HTML}{135579}
\definecolor{magenta}{HTML}{a8264f}

\newcommand{\commentstyle}{\color{dark-cyan}}
\newcommand{\errstyle}{\color{red}}
\newcommand{\lstcm}[1]{{\commentstyle#1}} %
\newcommand{\lsterrcm}[1]{{\errstyle#1}}

\lstdefinelanguage{LambdaCirc}%
{morekeywords={abstract,%
  case,catch,char,class,%
  def,else,extends,final,finally,for,%
  if,import,implicit,%
  match,module,%
  new,null,%
  object,override,%
  package,private,protected,public,%
  for,public,return,super,%
  this,throw,trait,try,type,%
  val,var,%
  with,while,%
  yield,%
  let,end,%
	in,fun,alloc,inc%
  },%
  mathescape=true,%
  sensitive,%
  keywordstyle={\color{dark-cyan}\bf\ttfamily},%
  commentstyle=\commentstyle,%
  morecomment=[l]//,%
  morecomment=[s]{/*}{*/},%
  morecomment=[s][\color{dark-cyan}]{@}{\ },%
  morestring=[b]",%
  morestring=[b]',%
  showstringspaces=false%
}[keywords,comments,strings]%

\newcommand{\trackset}[1]{^{\texttt{\{#1\}}}}

\newcommand{\trackvar}[1]{^{\texttt{#1}}}

\newcommand{\oldlang}{\ensuremath{\lambda^{\vardiamondsuit}}\xspace}
\newcommand{\maybelang}{\ensuremath{\mathsf{F}_{<:}^{\circ}}\xspace}
\newcommand{\natlang}{\ensuremath{\lambda_{\mathcal{N}}^{\circ}}\xspace}
\newcommand{\polylang}{\ensuremath{\mathsf{F}_{<:}^{\vardiamondsuit}}\xspace}

\makeatletter
\def\ifenv#1{%
   \def\@tempa{#1}%
   \ifx\@tempa\@currenvir
      \expandafter\@firstoftwo
    \else
      \expandafter\@secondoftwo
   \fi
}
\makeatother

\makeatletter
\edef\showenv{\@currenvir}
\makeatother

\newcommand{\Type}[1]{\ensuremath{\ifenv{lstlisting}{\texttt{#1}}{\mathsf{#1}}}}
\newcommand{\typevar}[1]{\ensuremath{\ifenv{lstlisting}{\texttt{#1}}{#1}}}
\newcommand{\ty}[2][]{\ensuremath{\ifthenelse{\isempty{#1}}{\typevar{#2}}{\typevar{#2}^{\,\typevar{#1}}}}}

\newcommand{\Var}{\Type{Var}}

\newcommand{\TRef}{\Type{Ref}}
\newcommand{\TTop}{\top}
\newcommand{\TUnit}{\Type{Unit}}
\newcommand{\Loc}{\Type{Loc}}

\newcommand{\TNat}{\Type{Nat}}
\newcommand{\TBool}{\Type{Bool}}

\newcommand{\tunit}{\Type{unit}}

\newcommand{\tref}{\Type{ref}}

\newcommand{\tsucc}[1]{\Type{succ}~#1}
\newcommand{\tpred}[1]{\Type{pred}~#1}
\newcommand{\tmul}[2]{\Type{mul}~#1~#2}
\newcommand{\tiszero}[1]{\Type{iszero}~#1}
\newcommand{\tif}[3]{\Type{if}~#1~\Type{then}~#2~\Type{else}~#3}
\newcommand{\ttrue}{\Type{true}}
\newcommand{\tfalse}{\Type{false}}
\newcommand{\tnat}[1]{\Type{nat}~#1}
\newcommand{\tbool}[1]{\Type{bool}~#1}

\newcommand{\ext}[1]{\HLBox[gray!20]{#1}}
\newcommand{\wrcolor}[1]{{\color{orange}{#1}}}
\newcommand{\rdcolor}[1]{{\color{blue}{#1}}}
\newcommand{\oldcolor}[1]{{\color{gray}{#1}}}

\newcommand{\SR}[2]{\TRef\ifenv{lstlisting}{}{~}[\ty[#2]{#1}]}
\newcommand{\SRf}[3]{\TRef\ifenv{lstlisting}{}{~}[\ty[\wrcolor{#2}\dots \rdcolor{#3}]{#1}]}
\newcommand{\SRef}[4]{\TRef\ifenv{lstlisting}{}{~}[\wrcolor{\ty[#2]{#1}}\dots\rdcolor{\ty[#4]{#3}}]}

\newcommand{\m}{\ensuremath{\bm{\mu}}}

\newcommand{\mty}[3][x]{\ensuremath{\m#1.\ty[#2]{#3}}}

\newcommand{\mr}[4][x]{\mty[#1]{#2}{\SR{#3}{#4}}}

\newcommand{\TAll}[6]{\ensuremath{\forall f(\ty[#2]{#1} <: \ty[#4]{#3}) . \ty[#6]{#5}}}
\newcommand{\TLam}[5]{\ensuremath{\Lambda f(\ty[#2]{#1}) . {#5}}}
\newcommand{\TApp}[3]{\ensuremath{{#1}\ [\ty[#3]{#2}]}}

\newcommand{\ts}[1][]{\ensuremath{\ifthenelse{\isempty{#1}}{\,\vdash\,}{\,\vdash^{\,#1}\,}}}

\newcommand{\flt}{\ensuremath{\varphi}}

\newcommand{\cx}[2][]{\ensuremath{\ifthenelse{\isempty{#1}}{#2}{#2^{\,#1}}}}

\providecommand{\G}{G} %
\renewcommand{\G}[1][]{\cx[#1]{\Gamma}}

\newcommand{\HLBox}[2][teal!12]{\ensuremath{\mathchoice%
  {\setlength{\fboxsep}{.5ex}\colorbox{#1}{$\displaystyle#2$}}%
  {\setlength{\fboxsep}{.5ex}\colorbox{#1}{$\textstyle#2$}}%
  {\setlength{\fboxsep}{.5ex}\colorbox{#1}{$\scriptstyle#2$}}%
  {\setlength{\fboxsep}{.5ex}\colorbox{#1}{$\scriptscriptstyle#2$}}}}%

\newcommand{\QFresh}{\ensuremath{\vardiamondsuit}}
\newcommand{\qbot}{\ensuremath{\varnothing}}
\newcommand{\qfresh}{\ensuremath{\vardiamondsuit}}
\newcommand{\subq}{\ensuremath{\subseteq}}
\newcommand{\qlub}{\ensuremath{\cup}}
\newcommand{\qglb}{\ensuremath{\cap}}

\xsavebox*{SMALLSTAR}{\(\raisebox{.25ex}{\(\qfresh\)}\)}

\xsavebox*{OVRLP}{$\raisebox{.37ex}[0pt][0pt]{$\mathrlap{\hspace{.415ex}\scaleobj{.5}{\vardiamondsuit}}$}\cap$}
\def\overlap{\ensuremath{\mathbin{\scalerel*{\xusebox{OVRLP}}{\sqcap}}}}

\newcommand{\WF}[1]{\ensuremath{#1\ \mathsf{ok}}}
\newcommand{\reaches}{\ensuremath{\mathrel{\leadsto}}}
\newcommand{\norm}[1]{\lvert #1 \rvert}
\newcommand{\cardinality}[2]{\norm{#1}_{#2}}
\newcommand{\qtrans}[2][]{\ensuremath{\ifthenelse{\isempty{#1}}{#2\mathord{*}}{{#2}^{#1}}}}
\newcommand{\sat}[1]{\mathsf{sat}\ #1}
\newcommand{\dsat}[1]{\mathsf{dsat}\ #1}
\newcommand{\psat}[1]{\mathsf{psat}\ #1}

\newcommand{\BOX}[1]{\fbox{$\strut #1$}}

\newcommand{\FV}{\ensuremath{\operatorname{fv}}}
\newcommand{\FTV}{\ensuremath{\operatorname{ftv}}}
\newcommand{\Singleton}{\ensuremath{\mathcal{P}_{1}}}

\newcounter{typerule}
\crefformat{typerule}{#2\textsc{#1}#3}
\Crefformat{typerule}{#2\textsc{#1}#3}

\newcommand{\typerule}[3]{%
  \def\thetyperule{#1}%
  \refstepcounter{typerule}%
  \label{typing:#1}%
  \infrule[#1]{#2}{#3}
}
\newcommand{\vgap}{\vspace{7pt}}

\newcommand{\hole}[1]{\ensuremath{[\,#1\,]}}
\newcommand{\CX}[3][black]{\ensuremath{{\color{#1}#2\ifthenelse{\isempty{#3}}{}{\hole{{\color{black}#3}}}}}}

\newcommand{\rulename}[1]{(\textsc{#1})} %

\graphicspath{ {./images/} }
\setcopyright{none}

\usepackage[normalem]{ulem}
\makeatletter
\UL@protected\def\reduwave{\leavevmode \bgroup
\ifdim \ULdepth=\maxdimen \ULdepth 3.5\p@
\else \advance\ULdepth2\p@
\fi \markoverwith{\lower\ULdepth\hbox{\textcolor{red}{\sixly \char58}}}\ULon}
\makeatother
\makeatletter%
\begin{document}

\title{Complete the Cycle: Reachability Types with Expressive Cyclic References (Extended Version)}

\author{Haotian Deng}
\orcid{0009-0002-7096-2646}
\affiliation{%
  \institution{Purdue University}
  \city{West Lafayette}
  \country{USA}
}
\email{deng254@purdue.edu}

\author{Siyuan He}
\orcid{0009-0002-7130-5592}
\affiliation{%
  \institution{Purdue University}
  \city{West Lafayette}
  \country{USA}
}
\email{he662@purdue.edu}

\author{Songlin Jia}
\orcid{0009-0008-2526-0438}
\affiliation{%
  \institution{Purdue University}
  \city{West Lafayette}
  \country{USA}
}
\email{jia137@purdue.edu}

\author{Yuyan Bao}
\orcid{0000-0002-3832-3134}
\affiliation{%
  \institution{Augusta University}
  \city{Augusta}
  \country{USA}
}
\email{yubao@augusta.edu}

\author{Tiark Rompf}
\orcid{0000-0002-2068-3238}
\affiliation{%
  \institution{Purdue University}
  \city{West Lafayette}
  \country{USA}
}
\email{tiark@purdue.edu}

\lstMakeShortInline[keywordstyle=,%
                    flexiblecolumns=false,%
                    language=Scala,
                    basewidth={0.56em, 0.52em},%
                    mathescape=false,%
                    basicstyle=\footnotesize\ttfamily]@

\begin{abstract}
Local reasoning about programs that combine aliasing and mutable state is a
longstanding challenge. Existing approaches -- ownership systems, linear and
affine types, uniqueness types, and lexical effect tracking -- impose global
restrictions such as uniqueness or linearity, or rely on shallow syntactic
analyses. These designs fall short with higher-order functions and shared
mutable state. Reachability Types (RT) track aliasing and separation in
higher-order programs, ensuring runtime safety and non-interference. However, RT
systems face three key limitations: (1) they prohibit cyclic references, ruling
out non-terminating computations and fixed-point combinators; (2) they require
deep tracking, where a qualifier must include all transitively reachable
locations, reducing precision and hindering optimizations like fine-grained
parallelism; and (3) referent qualifier invariance prevents referents from
escaping their allocation contexts, making reference factories inexpressible. 

In this work, we address these limitations by extending RT with three mechanisms
that enhance expressiveness. First, we introduce cyclic references, enabling
recursive patterns to be encoded directly through the store. Second, we adopt
shallow qualifier tracking, decoupling references from their transitively
reachable values. Finally, we introduce an escaping rule with reference
subtyping, allowing referent qualifiers to outlive their allocation context.
These extensions are formalized in the $\mathsf{F}_{<:}^{\circ}$-calculus with a
mechanized proof of type soundness, and case studies illustrate expressiveness
through fixpoint combinators, non-interfering parallelism, and escaping
read-only references. \end{abstract}

\maketitle

\section{Introduction}
\label{sec:intro}

Local reasoning in the presence of aliasing and mutable state is a foundational
challenge in programming languages. This challenge gave birth to an extensive
list of ideas, from ownership
systems~\Citep{DBLP:conf/oopsla/ClarkePN98,DBLP:conf/ecoop/ClarkeNP01,DBLP:series/lncs/ClarkeOSW13,
DBLP:conf/oopsla/Hogg91}, to linear/affine
types~\Citep{DBLP:conf/ifip2/Wadler90, DBLP:journals/tcs/Girard87}, to
uniqueness types~\Citep{DBLP:journals/mscs/BarendsenS96}, to capability-based
region type systems 
\Citep{ACMTrans.Program.Lang.Syst./WalkerK00,Proc.26thACMSIGPLAN-SIGACTSymp.Princ.Program.Lang./CraryD99}.

Not only have these ideas shaped theoretical developments, they also influenced
the design of mainstream programming languages. Notably,
Rust~\Citep{DBLP:conf/sigada/MatsakisK14} has emerged as a gold standard for
practical ownership-based systems, combining affine types and lexical lifetimes
to ensure strong memory and concurrency safety. However, Rust enforces a strict
aliasing discipline: its ``shared XOR mutable'' invariant requires global
uniqueness of mutable references. This restricts expressiveness in common idioms
involving shared mutable references, cyclic structures, or higher-order
functions that close over mutable state.

Reachability Types
(RT)~\citep{DBLP:journals/pacmpl/BaoWBJHR21,DBLP:journals/pacmpl/WeiBJBR24}  
offer a promising alternative to bring expressive, fine-grained reasoning about
resource sharing and separation to higher-level functional languages, drawing
inspiration from separation
logic~\citep{DBLP:conf/lics/Reynolds02,DBLP:conf/csl/OHearnRY01}. Existing RT
systems demonstrate partial success in flexible reasoning about safety and
separation, from key abstraction mechanisms such as higher-order functions
\citep{DBLP:journals/pacmpl/BaoWBJHR21} to mutable state and polymorphic types
\citep{DBLP:journals/pacmpl/WeiBJBR24}. For example, RT successfully type
challenging patterns involving closures that share mutable state -- patterns
beyond the reach of ownership, linearity, or lexical effect systems (See Figure
1 in \Citet{DBLP:journals/pacmpl/WeiBJBR24}).

However, the core of reference types in existing RT systems has received 
comparatively less attention and remains limited in several key aspects: (1)
They lack support for cyclic references and recursive functions 
(\Cref{sec:intro-cyclic}); (2) their qualifiers overapproximate by tracking all 
transitively reachable heap locations (\Cref{sec:intro-shallow}); (3) they 
disallow semantically valid reference escapes (\Cref{sec:intro-escape}).

\begin{figure}[tb]
\begin{mdframed}
\begin{minipage}[t]{1\textwidth}
\begin{subfigure}[t]{.48\textwidth}
\begin{lstlisting}[xleftmargin=.05\textwidth,numbers=left]
val c = ... // : Ref[(Unit => Unit)$\lstcm{\trackvar{q}}$]$\lstcm{\trackvar{c}}$
def f(x : Unit) = {(!c)(x)} 
// $\lstcm{\dashv}$ [ f : (Unit => Unit)$\lstcm{\trackvar{c}}$ ]
c := f // $\lstcm{\text{c} \not<: \text{q}}$
// $\lsterrcm{\text{Error! Referent qualifier mismatch}}$
\end{lstlisting}
\captionsetup{justification=raggedright,singlelinecheck=false}
\caption{
    Landin's knot is not typeable in \oldlang
    \citep{DBLP:journals/pacmpl/WeiBJBR24} due to the lack of cyclic
    references. Line 4 fails to type check because the referent qualifier
    \texttt{q} can only reach observable resources before \texttt{c} is
    declared, which does not include \texttt{c}.
}\label{fig:example-landin-diamond}
\end{subfigure}
\hfill
\vline
\hfill
\begin{subfigure}[t]{.48\textwidth}
\begin{lstlisting}[xleftmargin=.07\textwidth,numbers=left]
val c = ... // : $\lstcm{\mr[z]{c}{(\TUnit\ => \TUnit)}{z}}$
def f(x : Unit) = {(!c)(x)} 
// $\lstcm{\dashv}$ [ f : (Unit => Unit)$\lstcm{\trackvar{c}}$ ]
c := f
// Okay, because c is a cyclic reference
\end{lstlisting}
\captionsetup{justification=raggedright,singlelinecheck=false}
\caption{
    Landin's knot typed with a cyclic reference in \maybelang (this work). Line
    4 is type checked with \Cref{typing:t-sassgn-v} (\Cref{fig:maybe-typing}).
    Its reachability graph is shown in \Cref{fig:graph-cycle}.
}\label{fig:example-landin-circle}
\end{subfigure}
\end{minipage}

\caption{Typing Cyclic References (\Cref{sec:intro-cyclic}): not possible in
\Citet{DBLP:journals/pacmpl/WeiBJBR24} (\Cref{fig:example-landin-diamond}),
possible in this work
(\Cref{fig:example-landin-circle}).}\label{fig:example-landin}
\end{mdframed}
\vspace{-3ex}
\end{figure} %
\begin{figure}[tb]
\begin{mdframed}
\begin{minipage}[t]{\textwidth}
\begin{subfigure}[t]{.48\textwidth}
\begin{lstlisting}[xleftmargin=.05\textwidth,numbers=left]
val inner = new Ref(...) // : $\lstcm{\ty[inner]{\TRef[...]}}$
val outer = new Ref(inner) 
// : $\lstcm{\ty[outer,inner]{\TRef[\ty[inner]{\TRef[...]}]}}$
def par(b1: (Unit => Unit)$^{\QFresh}$)
       (b2: (Unit => Unit)$^{\QFresh}$) = ...
// parallelize b1 and b2
par { inner := ... } { outer := ... }
// $\lsterrcm{\texttt{Error! inner overlaps outer}}$
\end{lstlisting}
\captionsetup{justification=raggedright,singlelinecheck=false}
\caption{
    \oldlang's deep reference tracking always make reference track their
    referents. Line 7 fails to type check because \texttt{inner} is considered
    part of \texttt{outer}, making it impossible to parallelize uses of the two
    resources.
}
\label{fig:shallow-old}
\end{subfigure}
\hfill
\vline
\hfill
\begin{subfigure}[t]{.48\textwidth}
\begin{lstlisting}[xleftmargin=.07\textwidth,numbers=left]
val inner = new Ref(...) // $\lstcm{\ty[\texttt{inner}]{\TRef[...]}}$
val outer = new Ref(inner)
// : $\lstcm{\ty[outer]{\TRef[\ty[inner]{\TRef[...]}]}}$
def par(b1: (Unit => Unit)$^{\QFresh}$)
       (b2: (Unit => Unit)$^{\QFresh}$) = ...
// parallelize b1 and b2
par { inner := ... } { outer := ... }
// Okay, inner and outer are disjoint
\end{lstlisting}
\captionsetup{justification=raggedright,singlelinecheck=false}
\caption{
    Under shallow reference tracking in \maybelang (this work), line 7 is
    type checked because \texttt{inner} and \texttt{outer} are disjoint, thus
    can be safely parallelized.
}
\label{fig:shallow-new}
\end{subfigure}

\end{minipage}
\caption{Qualifier separation between reference and referent
(\Cref{sec:intro-shallow}): not possible in
\Citet{DBLP:journals/pacmpl/WeiBJBR24} (\Cref{fig:shallow-old}), possible in
this work (\Cref{fig:shallow-new}).}
\label{fig:shallow-example}
\end{mdframed}
\vspace{-3ex}
\end{figure}
\begin{figure}[tb]
\begin{mdframed}
\begin{minipage}[t]{\textwidth}
\begin{subfigure}[t]{.48\textwidth}
\begin{lstlisting}[xleftmargin=.05\textwidth,numbers=left]
def mkRef() =
    val x = ... // $\lstcm{\ty[c]{T}}$
    val c = new Ref(x) // $\lstcm{\ty[c]{\TRef[\ty[x]{T}]}}$
    c // $\lstcm{\ty[c]{\TRef[\ty[x]{T}]} \not<: \ty[c]{\TRef[\ty[\qbot]{T}]}}$
// $\lsterrcm{\text{Error! x is untracked outside mkRef}}$
\end{lstlisting}
\captionsetup{justification=raggedright,singlelinecheck=false}
\caption{
Reference factory functions not typeable in \oldlang. Line 4 fails because the
returning reference \texttt{c}'s referent contains local variable \texttt{x}
which cannot escape the function scope.
}\label{fig:escape-old}
\end{subfigure}
\hfill
\vline
\hfill
\begin{subfigure}[t]{.48\textwidth}
\begin{lstlisting}[xleftmargin=.07\textwidth,numbers=left]
def mkRef() =
    val x = ... // $\lstcm{\ty[c]{T}}$
    val c = new Ref(x) // $\lstcm{\ty[c]{\TRef[\ty[x]{T}]}}$
    c // Okay, $\lstcm{\ty[c]{\TRef[\ty[x]{T}]}}$ <: $\lstcm{\mr[z]{c}{T}{z}}$
mkRef(...) // Okay, $\lstcm{\mr[z]{\QFresh}{T}{\&z}}$
\end{lstlisting}
\captionsetup{justification=raggedright,singlelinecheck=false}
\caption{
A reference factory function typeable in \maybelang (this work). Line 4 is
successfully type checked by \Cref{typing:t-esc}, escaping to a read-only cyclic
reference (\Cref{fig:dual-component-semantics}).
}\label{fig:escape-new}
\end{subfigure}
\end{minipage}

\caption{Typing escaping references (\Cref{sec:intro-escape}): not possible in
\Citet{DBLP:journals/pacmpl/WeiBJBR24} (\Cref{fig:escape-old}), possible in this work
(\Cref{fig:escape-new}).}
\label{fig:escape-example}
\end{mdframed}
\vspace{-3ex}
\end{figure}

 \subsection{Cyclic References}\label{sec:intro-cyclic}

Recent theoretical advances in RT have shown that variants of RT that support
higher-order references remain \textit{terminating} \citep{bao_modeling_2023}.
This is surprising, because languages that combine higher-order functions with
higher-order mutable references typically permit general recursion by encoding
fixed-point combinators through the store. The key pattern underlying such
encodings is Landin's Knot \citep{TheComputerJournal/Landin64}, where a function
is stored in a reference that in turn is captured by the function itself,
creating a cycle in the store, thereby enabling non-terminating computations.

A typical encoding of Landin's Knot is shown in
\Cref{fig:example-landin-diamond}. The program creates a mutable reference @c@,
which initially stores a function value. The reference is updated with function
@f@ (at line 4), which captures @c@ and calls the function stored at reference
@c@. Calling the function stored at @c@ will recursively invoke the function
itself through the store, leading to infinite recursion.

However, the code in \Cref{fig:example-landin-diamond} is not typeable in
~\citep{DBLP:journals/pacmpl/WeiBJBR24}'s system, because in their system,
referent qualifiers are invariant: a reference can only be updated with values
having an \textit{equivalent} qualifier. In this example, when @c@ is
allocated, its type must be assigned before @c@ itself is in scope; and
the initial qualifier @q@ used in its type cannot include @c@,
since @c@ has not yet been bound. But the reference @c@ is updated with a
function that captures @c@ (at line 4), violating referent qualifier invariance.
As a result, general recursion via Landin's Knot is disallowed in their system.

In this work, we introduce \maybelang, an extension of
\citep{DBLP:journals/pacmpl/WeiBJBR24}'s $\polylang$ system, which supports
explicit cyclic reference types. As shown in~\Cref{fig:example-landin-circle},
variable @c@ now has a cyclic reference type. The $\mu z$ binder can be included
in the inner referent qualifier to refer to the outer reference itself, making
Landin's knot typeable in the system.

Extending prior work~\cite{DBLP:journals/pacmpl/WeiBJBR24} to support cyclic  
reference types introduces unique challenges: as we discuss in
\Cref{sec:motivation-cyclic}, a naive assignment rule for cyclic references is
unsound.  To preserve soundness, we impose two constraints on cyclic
assignments: (1) the assigned reference term must be a variable, and (2) the
assignee's qualifier must be a singleton matching that variable. A second
challenge is the need to adjust the application rule to ensure that cyclic
assignments in an abstraction are still well-typed after $\beta$-reduction.

We describe our approach to overcoming these challenges in
\Cref{sec:motivation-cyclic}, and present a general fixpoint operator as case
study in \Cref{sec:casestudy-fixpoint}.

\subsection{Flexible \& Precise Reference Qualifiers}\label{sec:intro-shallow}

\begin{figure}
    \begin{mdframed}
        \begin{subfigure}[t]{.95\textwidth}
            \begin{minipage}[t]{.4\linewidth}\vspace{0pt}
                \begin{tikzpicture}[node distance = 1.2cm]
                    \node[untracked lambda, draw, aspect=2.5] (c) {$\lambda$} node[below=.5cm] (lc) {\texttt{T}$^\bot$};

                    \node[locs, draw, below left of=c] (n1) {} node[below=.05cm of n1] (ln1) {\texttt{Ref[T$^\bot$]}};
                    \node[locs, draw, below right of=c] (n2) {} node[below=.05cm of n2] (ln2) {\texttt{Ref[T$^\bot$]}};

                    \draw[nodearrow] (n1) -- (c);
                    \draw[nodearrow] (n2) -- (c);
                \end{tikzpicture}
            \end{minipage}
            \begin{minipage}[t]{.6\linewidth}\vspace{0pt}
                \caption{Reachability graph of
                \citet{DBLP:journals/pacmpl/BaoWBJHR21}'s system. Because
                references are first-order with untracked referents (depicted as
                a dotted \colorbox{gray!10}{gray} cloud), the system cannot
                represent hierarchical structures using reference types.}
            \end{minipage}
            \label{fig:graph-bao}
        \end{subfigure}

        \begin{subfigure}[t]{.48\textwidth}
            \centering
            \begin{tikzpicture}[node distance = 1.2cm]
                    \node[locs] (n11) at ([xshift=-1cm]current page.south) {}; \node[below=.05cm of n11] {\texttt{Ref[T$^\texttt{p}$]$^\QFresh$}};
                    \node[locs] (n12) at ([xshift=1cm]current page.south) {}; \node[below=.05cm of n12] {\texttt{Ref[T$^\texttt{q}$]$^\QFresh$}};

                    \node[above left of = n11] (n21) {...};
                    \node[locs, above of = n11] (n22) {};
                    \node[locs, above right of = n11] (n23) {};
                    \node[tracked lambda, above right of = n12] (n24) {$\lambda$};

                    \node[locs, above left of = n22] (n31) {};
                    \node[locs, above of = n22, xshift=.1cm] (n32) {};
                    \node[locs, above right of = n22] (n33) {};
                    \node[locs, above left of = n24] (n34) {};
                    \node[locs, above right of = n24] (n35) {};

                    \node[above of = n31, xshift=-.5cm] (n41) {...};
                    \node[untracked lambda, above of = n32] (n42) {$\lambda$};
                    \node[above right of = n32] (n43) {...};

                    \node[locs, above left of = n35] (n44) {};
                    \node[above of = n35] (n45) {...};

                    \node[above left of = n44] (n51) {...};
                    \node[above right of = n44] (n52) {...};

                    \draw[nodearrow] (n11) -- (n21);
                    \draw[nodearrow] (n11) -- (n22);
                    \draw[nodearrow] (n11) -- (n23);
                    \draw[nodearrow] (n12) -- (n24);

                    \draw[nodearrow] (n22) -- (n31);
                    \draw[nodearrow] (n22) -- (n32);
                    \draw[nodearrow] (n22) -- (n33);
                    \draw[nodearrow] (n24) -- (n34);
                    \draw[nodearrow] (n24) -- (n35);

                    \draw[nodearrow] (n31) -- (n41);
                    \draw[nodearrow] (n31) -- (n42);
                    \draw[nodearrow] (n32) -- (n42);
                    \draw[nodearrow] (n33) -- (n43);
                    \draw[nodearrow] (n34) -- (n44);
                    \draw[nodearrow] (n35) -- (n44);
                    \draw[nodearrow] (n35) -- (n45);

                    \draw[nodearrow] (n44) -- (n51);
                    \draw[nodearrow] (n44) -- (n52);
            \end{tikzpicture}
            \caption{Reachability graph illustrating the ``deep separation''
            reference tracking model in \Citet{DBLP:journals/pacmpl/WeiBJBR24}'s
            system.  References marked with \QFresh\ are separate from each
            other and enforce strict separation of all transitively reachable
            locations.  Untracked data, which separate references may still
            reach, is depicted as dotted \colorbox{gray!10}{gray} clouds.}
            \label{fig:graph-wei}
        \end{subfigure}
        \hfill
        \vline
        \hfill
        \begin{subfigure}[t]{.48\textwidth}
            \centering
            \begin{tikzpicture}[node distance = 1.2cm]
                    \node[locs] (n11) at ([xshift=-1cm]current page.south) {}; \node[below=.05cm of n11] {\texttt{Ref[T$^\texttt{p}$]$^\QFresh$}};
                    \node[locs] (n12) at ([xshift=1cm]current page.south) {}; \node[below=.05cm of n12] {\texttt{Ref[T$^\texttt{q}$]$^\QFresh$}};

                    \node[above left of = n11] (n21) {...};
                    \node[locs, above of = n11] (n22) {};
                    \node[shared locs, above right of = n11] (n23) {};
                    \node[tracked lambda, above right of = n12] (n24) {$\lambda$};

                    \node[locs, above left of = n22] (n31) {};
                    \node[locs, above of = n22, xshift=.1cm] (n32) {};
                    \node[shared locs, above right of = n22] (n33) {};
                    \node[locs, above left of = n24] (n34) {};
                    \node[circular locs, above right of = n24] (n35) {};

                    \node[above of = n31, xshift=-.5cm] (n41) {...};
                    \node[shared lambda, above of = n32] (n42) {$\lambda$};
                    \node[shared locs, above right of = n33] (n43) {};
                    \node[tracked lambda, above of = n35, xshift=-.8cm] (n44) {$\lambda$};

                    \node[above right of = n42] (n51) {...};
                    \node[above left of = n43, xshift=.5cm] (n52) {...};
                    \node[above right of = n43] (n53) {...};

                    \draw[nodearrow] (n11) -- (n21);
                    \draw[nodearrow] (n11) -- (n22);
                    \draw[nodearrow] (n11) -- (n23);
                    \draw[nodearrow] (n12) -- (n24);
                    \draw[nodearrow] (n12) -- (n23);

                    \draw[nodearrow] (n22) -- (n31);
                    \draw[nodearrow] (n22) -- (n32);
                    \draw[nodearrow] (n22) -- (n33);
                    \draw[nodearrow] (n23) -- (n33);
                    \draw[nodearrow] (n24) -- (n34);
                    \draw[nodearrow] (n24) -- (n35);

                    \draw[nodearrow] (n31) -- (n41);
                    \draw[nodearrow] (n31) -- (n42);
                    \draw[nodearrow] (n32) -- (n42);
                    \draw[nodearrow] (n33) -- (n43);
                    \draw[nodearrow] (n34) -- (n43);
                    \draw[nodearrow, bend left=60] (n35) to (n44);

                    \draw[nodearrow] (n42) -- (n51);
                    \draw[nodearrow] (n43) -- (n52);
                    \draw[nodearrow] (n43) -- (n53);
                    \draw[nodearrow, bend left=60] (n44) to (n35);
            \end{tikzpicture}
            \caption{Reachability graph illustrating the ``shallow separation''
            reference tracking model in \maybelang\ (this work). Unlike
            \citet{DBLP:journals/pacmpl/WeiBJBR24}'s system, our model permits
            separate references to share child nodes (depicted in
            \colorbox{yellow!20}{yellow}) and supports cyclic references where
            values can reach the references themselves (highlighted in
            \colorbox{purple!10}{purple}). }
            \label{fig:graph-this}
        \end{subfigure}
        \caption{Reachability graph illustrating references in
        \Citet{DBLP:journals/pacmpl/BaoWBJHR21},
        \Citet{DBLP:journals/pacmpl/WeiBJBR24}, and this work. The
        $\lambda$-labeled clouds represent non-location values, and the circular
        nodes ($\bigcirc$) represent store locations. All store locations are
        tracked (we denote tracked values using \colorbox{blue!10}{pale blue}).
        We discuss a concrete example in \Cref{sec:motivation-shallow}. Despite
        relaxing certain internal invariants, the end-to-end guarantees remain
        intact (See \Cref{sec:formal-meta}).
        }
        \label{fig:graph-tree}
    \end{mdframed}
    \vspace{-3ex}
\end{figure} %
 \Citet{DBLP:journals/pacmpl/WeiBJBR24}'s reference typing
employs a ``deep'' tracking mechanism to enforce strong separation between
portions of the heap. However, this limits the shape of heap structures that can
be formed using nested references. As illustrated in~\Cref{fig:graph-wei},
separate references\footnote{Separation is indicated by the freshness marker
$\QFresh$, which signifies contextual freshness. Resources marked with \QFresh\
reach contextually fresh objects, making them separate from other currently
observable resources.  See~\Cref{sec:motivation-background} for a detailed
explanation of the \QFresh\ marker.} remain distinct from each other in terms of
their \textit{transitively reachable locations}.\footnote{This follows directly from
\Cref{typing:t-ref}, where \texttt{q} is included in the qualifiers of both the
inner referent and the outer reference. As a result, the outer qualifier always
includes all transitively reachable locations.} This restriction prevents parent
references from sharing store locations, further restricting the flexibility of
heap structures with nested references.

The example of \Cref{fig:shallow-old} creates reference @outer@ that takes
@inner@ as its referent, and it is obvious that reference @outer@ resides in a
separate location from its referent @inner@. However, in their system, @outer@'s
qualifier surprisingly includes @inner@ (at line 3), making it impossible to
treat a reference and its underlying value as distinct entities. Consequently,
this deep tracking model forces unnecessary
dependencies~\cite{DBLP:journals/pacmpl/BracevacWJAJBR23}, hindering
optimizations such as parallelization (\eg, at line 8).  To address these
limitations, we refine the reference introduction rule to adopt a ``shallow''
reference tracking, where references and their referents remain distinct by
default. In this refined system, @inner@ and @outer@ remain separate, enabling
safe parallelization, as shown in \Cref{fig:shallow-new}. This also allows
distinct references to share common underlying objects, as shown in
\Cref{fig:graph-this}.

Adopting shallow reference tracking is non-trivial, as it weakens a key internal
invariant of prior RT systems, namely that each reference qualifier
conservatively captures all heap locations transitively reachable from its
referent. With shallow tracking, distinct qualifiers may now reach overlapping
regions of the heap, potentially breaking the separation guarantees of the
system.

To reestablish separation guarantees, we develop new internal invariants to
compensate for the loss of transitive reachability, demonstrating that the key
reference typing rules -- for assignment and dereferencing -- are strong enough
to enforce separation. In particular, we prove that the progress and
preservation in parallel reduction corollary (\Cref{sec:formal-static}) of
\citet{DBLP:journals/pacmpl/WeiBJBR24}'s system remain valid, even without
explicitly tracking transitively reachable locations.

We illustrate the expressiveness of this design with additional examples in
\Cref{sec:motivation-shallow} and a case study on fine-grained parallelism in
\Cref{sec:casestudy-shallow}.

\subsection{Escaping References}\label{sec:intro-escape}

Sometimes, the precision gained by the shallow model is undesirable, and a
\textit{controlled relaxation/imprecision} in reference tracking is beneficial,
particularly when a mutable reference escapes its defining scope.
Interestingly, \Citet{DBLP:journals/pacmpl/WeiBJBR24} does not allow escaping 
references that capture resources declared in the escaped scope. As shown in
\Cref{fig:escape-old}, function @mkRef@ attempts to return a reference to
@x@, but since @x@ is a local variable, the qualifier rules in
their system prohibit it from escaping, leading to a type error. This suggests
that the system is too strict in its reference tracking, and that a more
flexible system is needed to support escaping references.

A key challenge in supporting escaping references is that referent types 
are \emph{invariant}:
references cannot be covariant (allowing writing larger-than-expected values)
nor contravariant (allowing reading smaller-than-expected values). To address
this limitation, we introduce dual-component references of the form
$\mty{\QFresh}{(\SRef{T}{x,q}{U}{x,p})}$ with separate
\wrcolor{write} and \rdcolor{read} components (see
\Cref{fig:dual-component-semantics}). This formulation sidesteps the invariance
challenge by splitting the referent's qualifier into two parts -- intuitively,
allowing us to ``put in'' less while ``getting out'' more, thus maintaining
soundness. 

Dual-component references enable controlled reference escaping
(\Cref{sec:dual}). In \Cref{fig:escape-new}, before reference @c@ goes out of
scope, we upcast its type to the read-only cyclic reference type, \ie,
$\mr[z]{c,x}{T}{\&z}$, through reference subtyping with an escaping rule.  The
escaping rule ensures that the \rdcolor{read qualifier} of @x@ is
transferred to an outer scope, enabling the function to return @c@ with
an extended, yet sound, reachability qualifier (see \Cref{typing:t-esc} in
\Cref{fig:dual-component-semantics}). This extension allows examples such as
reference factory functions to be type checked.

Beyond escaping, dual-component references also enable general subtyping of
referents (see \Cref{typing:s-sref} in \Cref{fig:dual-component-semantics}),
allowing for finer-grained resources control, such as read-only references. In
\Cref{sec:casestudy-escape}, we demonstrate how dual-component references with
the escape rule enable new patterns of reference cell usage, further extending
the expressiveness of the type system.

\subsection{Contributions and Organization}

In summary, the main contributions of this paper are as follows:

\begin{itemize}[leftmargin=2em]
  
  \item After providing an overview of RT, we identify limitations in prior
  works, informally introduce the $\maybelang$-calculus, highlight key use cases
  and demonstrate expressiveness (\Cref{sec:motivation}).

 \item We present the formal theory and metatheory of the \maybelang-calculus, a
 variant of the RT systems that permits well-typed cyclic structures with a
 cyclic reference type and refined qualifier tracking mechanism
 (\Cref{sec:formal}).

  \item We extend \maybelang with dual-component references and an escaping rule
  that allows referent qualifiers to extend beyond their defining scope while
  maintaining soundness (\Cref{sec:dual}), enabling controlled imprecision in
  reference tracking.

  \item We demonstrate \maybelang's ability to handle well-typed cyclic
  references for store-based recursion (\Cref{sec:casestudy-fixpoint}), shallow
  reference tracking for fine-grained parallelism
  (\Cref{sec:casestudy-shallow}), and escaping references for reference factory
  functions (\Cref{sec:casestudy-escape}). These case studies showcase its
  improved precision and expressiveness over prior systems.

\end{itemize}

We address limitations of this work as well as the trade-offs that justify our
design decisions in \Cref{sec:limitation}, and discuss related work in
\Cref{sec:related}. The formal results in this paper have all been mechanized in
Rocq. All key examples from the paper have been mechanized; the remaining
examples can be directly derived from these. They are presented using our Rocq
term syntax with a standard locally nameless representation
\Citep{JAutomReasoning/Chargueraud12}. The development is available online
at \url{https://github.com/tiarkrompf/reachability}.

\section{Reachability Types (RT)} \label{sec:motivation} 
In this section, we review the key concepts in \citet{DBLP:journals/pacmpl/WeiBJBR24}'s system 
(\Cref{sec:motivation-background}), and informally introduce cyclic references and their role in
typing examples such as Landin's Knot (\Cref{sec:motivation-cyclic}), 
shallow reference tracking (\Cref{sec:motivation-shallow}), and 
escaping references (\Cref{sec:motivation-escape}). 

\subsection{Key Ideas of RT}
\label{sec:motivation-background}

\paragraph{\textbf{Reachability Qualifiers: Tracking Reachable Resources in Types}}

Types in RT are of the form $\ty[p]{T}$, where $p$ is a \textit{reachability
qualifier}, indicating the set of locations that may be reached from the result
of an expression. Reachability qualifiers are finite sets of variables, which
may optionally include the freshness marker $\QFresh$, indicating a fresh,
unnamed resource. In the following example, evaluating the expression @new Ref(0)@
results in a \emph{fresh} value: it is not yet bound to a name, but must be
tracked. The typing context \lstinline|$\lstcm{\text{[ counter: Ref[Int]}^{\QFresh} ]}$|
following a reverse turnstile "$\dashv$" means @counter@ reaches a fresh value:

\begin{lstlisting}
  val counter = new Ref(0)        // : Ref[Int]$^\lstcm{\trackvar{counter}}$ $\lstcm{\dashv}$ [ counter: Ref[Int]$^\lstcm{\QFresh}$ ]
\end{lstlisting}

RT keep reachability sets minimal, \eg, variable @counter@ tracks exactly
itself.  When an alias is created for variable @counter@ as shown below, RT
assign the one-step reachability set @counter2@ to variable @counter2@. 

\begin{lstlisting}
  val counter2 = counter          // : Ref[Int]$^\lstcm{\trackvar{counter2}}$ $\lstcm{\dashv}$ [ counter2: Ref[Int]$^\lstcm{\trackvar{counter}}$, counter: Ref[Int]$^\lstcm{\QFresh}$ ]
\end{lstlisting}

\noindent We can retrieve the complete reachability set by computing its
transitive closure with respect to the typing
context~\cite{DBLP:journals/pacmpl/WeiBJBR24}.

Functions also track reachability: their reachability qualifier includes all
\textit{captured variables}. In the following example, function @inc@ captures
the free variable @counter@:

\begin{lstlisting}
  def inc(n: Int) = { counter := !counter + n } // : inc: (Int => Unit)$\lstcm{\trackvar{counter}}$ $\lstcm{\dashv}$ [ counter: Ref[Int]$^\lstcm{\QFresh}$ ]
\end{lstlisting}

\vspace{-4pt}
\paragraph{\textbf{Freshness Marker in Function Arguments: Contextual Freshness}}

The presence of the freshness marker \QFresh\ in a function argument's qualifier
indicates that the argument may only reach \textit{unobservable} resources,
meaning that its reachable locations must remain separate from those of the
function. Thus, function applications must satisfy the \textit{separation
constraint}, requiring that the argument's reachability qualifier is disjoint
from that of the function.

\begin{lstlisting}
  def id(x: T$^\QFresh$): T$\trackvar{x}$ = x                                       // : ((x: T$^\lstcm{\QFresh}$) => T$^\lstcm{\trackset{x}}$)$^\lstcm{\qbot}$
\end{lstlisting}

\noindent The type means that function @id@ cannot capture anything from its context, and 
it accepts arguments that may reach unobservable resources. A function application that violates
this separation constraint results in a type error:

\begin{lstlisting}
  def update(x: Ref[Int]$^\QFresh$): Unit = { counter := !(x) + 1 }   // : ((x: Ref[Int]$^\lstcm{\QFresh}$) => Unit)$^\lstcm{\trackvar{counter}}$
  update(counter) // $\lsterrcm{\text{Error! variable counter overlaps with function update}}$
\end{lstlisting}

\noindent The above function application violates the
separation constraint: the passing argument @counter@ overlaps with function @update@.

\paragraph{\textbf{Reference Type: Mutable Cells}}
\label{sec:motivation-background-reference}

So far, the reference type examples have only used primitive referent types (\eg
@Int@) as the referent type. Since these have untracked qualifiers
($\qbot$)\footnote{The untracked qualifier ($\qbot$) indicates that a value has
no reachable locations. Primitive values are usually untracked, since they
represent pure, location-independent data.}, the qualifiers are elided.
\Citet{DBLP:journals/pacmpl/WeiBJBR24}'s system supports reference types with
tracked referent qualifiers, enforcing that an assigned referent must match the
exact specified qualifier:

\begin{minipage}{0.5\linewidth}
\begin{lstlisting}
val a = ...             // : T$\lstcm{\trackvar{a}}$  
val b = ...             // : T$\lstcm{\trackvar{b}}$
val cell = new Ref(...) // : Ref[T$^\lstcm{\trackvar{a}}$]$^\lstcm{...}$
cell := a // Okay
cell := b // $\lsterrcm{\text{Error! Referent qualifier mismatch!}}$
\end{lstlisting}%
\end{minipage}%
\begin{minipage}{0.5\linewidth}
\begin{lstlisting}
val a = ...             // : T$\lstcm{\trackvar{a}}$
val b = ...             // : T$\lstcm{\trackvar{b}}$
val cell = new Ref(...) // : Ref[T$^\lstcm{\trackvar{a,b}}$]$^\lstcm{...}$
cell := a // Okay, T$^\lstcm{\trackvar{a}}$ <: T$^\lstcm{\trackvar{a,b}}$
cell := b // Okay, T$^\lstcm{\trackvar{b}}$ <: T$^\lstcm{\trackvar{a,b}}$
\end{lstlisting}
\end{minipage}%

\noindent As shown above (left), since reference @cell@ has @a@ as its referent
qualifier, it is only permitted to be assigned a value with qualifier @a@.
Assigning it with a different qualifier, \eg, @b@, results in a type error.

Qualifiers in RT can be widened to a ``larger'' qualifier by the qualifier
subsumption rule (see \Cref{typing:sq-sub} in \Cref{fig:maybe-subtyping}). For
example, in the code above (right), the qualifier of the reference @cell@ is
adjusted to allow both @a@ and @b@ via qualifier subtyping.  Since both
\texttt{T}$^{\trackvar{a}}$ and \texttt{T}$^{\trackvar{b}}$ are subtypes of
\texttt{T}$^{\trackvar{a,b}}$, @cell@ may safely be assigned either.

\begin{figure}
\begin{tabularx}{\textwidth}{|p{0.3\textwidth}|>{\hsize=0.4\hsize}X|>{\hsize=1.6\hsize}X|>{\hsize=0.5\hsize}X|>{\hsize=0.5\hsize}X|}
    \hline
    & \texttt{e}
    & \texttt{y = new Ref(e)} 
    & \texttt{! y} 
    & \texttt{y := x} \\ 
    \hline
    
    First-Order Reference\newline (\Citet{DBLP:journals/pacmpl/BaoWBJHR21})\multifootnotemark[l1] 
    & $e : \ty[\bot]{B}$
    & $y: \ty[{q}]{\TRef~[\ty[\bot]{B}]}$
    & $!y : \ty[\bot]{B}$ 
    & $x : \ty[\bot]{B}$ \\ 
    \hline
    
    Nested Reference (Deep) \newline (\Citet{DBLP:journals/pacmpl/WeiBJBR24})\multifootnotemark[l2] 
    & $e : \ty[q]{T}$
    & $y: \ty[{q,\QFresh}]{\TRef~[\ty[q]{T}]}$
    & $!y : \ty[q]{T}$ 
    & $x : \ty[q]{T}$ \\ 
    \hline
    
    Nested Reference (Shallow) \newline (\Cref{sec:formal} in this work)
    & $e : \ty[q]{T}$ 
    & $y: \ty[{\QFresh}]{\TRef~[\ty[q]{T}]}$
    & $!y : \ty[q]{T}$
    & $x : \ty[q]{T}$ \\ 
    \hline
    
    Cyclic Reference\newline (\Cref{sec:formal} in this work) 
    & $e : \ty[q]{T}$ 
    & $y: \mr{\QFresh}{T}{x,q}$
    & $!y : \ty[y,q]{T}$
    & $x : \ty[y,q]{T}$ \\ 
    \hline
    
    Dual-Component Reference\newline (\Cref{sec:dual} in this work) 
    & $e : \ty[q]{T}$
    & $y: \mty{\QFresh}{(\SRef{T}{x,q}{U}{x,p})}$
    & $!y : \rdcolor{\ty[y,p]{U}}$
    & $x: \wrcolor{\ty[y,q]{T}}$ \\
	\hline
\end{tabularx}
\caption{Comparison of reference typing rules in RT. Prior works only support
first-order reference typing (row 1) and nested references with transitive/deep
qualifier tracking (rows 2).  This work introduces shallow reference typing (row
3), cyclic reference typing (row 4), and dual-component reference typing (row
5).} 
\label{fig:comparison}
\vspace{-3ex}
\end{figure}

\multifootnotetext[l1]{\Citet{DBLP:journals/pacmpl/BaoWBJHR21} used $\bot$ to
denote untracked qualifiers, which is semantically equivalent to $\qbot$ in
later systems. Their system does not have an equivalent concept of
``freshness''. Additionally, their system uses first-order references, where
referents can only have untracked primitive/base types, as denoted by $B$ (See
\Cref{fig:maybe-syntax}).}

\multifootnotetext[l2]{In this system, it is possible for references to have the
$\QFresh$ qualifier alone under a different context through fresh application
(\Cref{typing:t-app-fresh}), but it is impossible to produce separate references
that encapsulate a shared value (See \Cref{sec:motivation-shallow}).}
 
\paragraph{\textbf{Limited Reference Typing}} As shown in
~\Cref{fig:comparison}, \citet{DBLP:journals/pacmpl/BaoWBJHR21} introduced RT 
that support only untracked, first-order references, where references can only
refer to untracked primitive values. This means that references in their system
cannot contain functions or references, since they are always tracked by the
systems. \citet{DBLP:journals/pacmpl/WeiBJBR24} extended their work to support
nested references, where references can enclose other references. However, their
system (1) lacks support for cyclic references; (2) employs imprecise aliasing
tracking; (3) does not support escaping references. In the following three
sections, we informally introduce our $\maybelang{}$ calculus that effectively
addresses the three limitations respectively.

\subsection{Cyclic References}
\label{sec:motivation-cyclic}

We introduce cyclic reference type that is of the form $\mr[z]{p}{T}{q}$, where
$\m z$ creates a binder $z$ that binds the outer qualifier $p$, enabling the
referent qualifier $q$ to reference its own enclosing reference. Consider the
following example:

\begin{lstlisting}
  val outer = new Ref(...) // : $\lstcm{\mr[z]{outer}{T}{z}}$ $\lstcm{\dashv}$ [ outer: $\lstcm{\mr[z]{\QFresh}{T}{z}}$ ]
\end{lstlisting}

\noindent The code declares reference @outer@ with a cyclic reference type, where the
bound variable $z$ appears in the referent qualifier. It permits assignments
with values referencing itself:

\begin{lstlisting}
  val inner = ...          // inner : $\lstcm{\ty[outer]{T}}$
  outer := inner           // Okay, because outer is a cyclic reference
\end{lstlisting}

\noindent This construct introduces a cyclic dependency: variable @inner@ 
\textit{reaches} reference @outer@ (as indicated by its qualifier), while
reference @outer@, being a cyclic reference, can \textit{contain} @inner@
via the assignment.

However, soundly incorporating cyclic dependencies into the type system is
non-trivial. Consider the following naive (and erroneous) assignment rule
\Cref{typing:t-sassgn-err}: \footnote{This rule also lacks an additional premise
$\QFresh \notin q$, which is not discussed in this section. The design decisions
around this rule are explored further in \Cref{sec:formal-static}.}

\vspace{-2ex}
\begin{minipage}[t]{.9\linewidth}\small
\typerule{t-sassgn-err}{
    \G[\flt]\ts t_1 : \mr[z]{p}{T}{\ext{z,q}} \qquad
    \G[\flt]\ts t_2 : \ty[\ext{p,q}]{T}
}{
    \G[\flt]\ts t_1 \coloneqq t_2 : \ty[\qbot]{\TUnit}
}
\end{minipage}

\noindent where the first premise requires that $t_1$ has a cyclic reference
type, with its referent qualifier possibly reaching either $q$ or itself
(indicated by $z$). The second premise states that any term $t_2$ can be
assigned to $t_1$, as long as its qualifier does not exceed $q$ and the
qualifier of $t_1$, denoted by $p$.

While \Cref{typing:t-sassgn-err} provides flexibility for cyclic dependencies, 
it permits an assignee with \textit{any} qualifier to be assigned to a
self-reference through subtyping, violating the intended semantics of cyclic
references.  Consider the following example: 

\begin{lstlisting}
  val e1 = new Ref(...)    // : $\lstcm{\mr[x]{e1}{T}{x}}$
  val e2 = ...             // : $\lstcm{\ty[e2,a,b,c,d,e]{T}}$
  e1 := e2                 // $\lsterrcm{\texttt{Unsound operation permitted by t-sassgn-err!}}$
  // e1: $\lstcm{\mr[x]{e1}{T}{x}}$    <: $\lstcm{\mr[x]{e1,e2,a,b,c,d,e}{T}{x}}$, e2: $\lstcm{\ty[e2,a,b,c,d,e]{T}}$ <: $\lstcm{\ty[e1,e2,a,b,c,d,e]{T}}$
\end{lstlisting}

\noindent where the qualifier of the cyclic reference @e1@
is incorrectly widened to accommodate an arbitrary qualifier in @e2@, while @e2@
is widened to include @e1@ in its qualifier.
This is clearly unsound, yet it would be naively permitted by 
\Cref{typing:t-sassgn-err}. This issue arises because upcasting of a cyclic
reference's outer qualifier allows it to generalize beyond its original scope.
Thus, if assignments were based on the outer \textit{qualifier} (as in
\Cref{typing:t-sassgn-err}), the system's qualifier tracking would become
inconsistent.

To prevent this unsound behavior, we restrict assignments so that the assigned
reference must have the form of a \textit{variable binder}, and the assignee's qualifier must be
a \textit{singleton qualifier}\footnote{We introduce singleton qualifiers to
make rules for cyclic references concise. For example, in
\Cref{typing:t-sassgn-v} (\Cref{fig:maybe-typing}), we require the assignee of a
cyclic reference to contain precisely a singleton variable qualifier matching
that of the reference. Under static semantics, a singleton qualifier contains
precisely a single variable (see \Cref{fig:maybe-syntax}) and under dynamic
semantics, it may also be a single location (see \Cref{fig:maybe-semantics}).}
matching that variable.

Restricting cyclic assignment to variable binders and singleton qualifiers is
consistent with the reduction semantics (see \Cref{fig:maybe-semantics}), as a variable of reference type (\eg, @e1@
in this case) will always be substituted into a single location, and the
assignee's singleton variable qualifier (\eg, @e2@) will be substituted into the
exact same singleton location qualifier (see \Cref{sec:typing-assgn}).

With the correct typing rule in place, the system would reject the invalid
assignment operation in the previous example, thereby retaining sound qualifier
tracking:

\begin{lstlisting}
    e1 := e2 // $\lsterrcm{\text{Error! e2's qualifier must be exactly e1}}$
    // e2: $\lstcm{\ty[e2,a,b,c,d,e]{T}}$ $\not$<: $\lstcm{\ty[e1]{T}}$
\end{lstlisting}

\paragraph{\textbf{Landin's Knot: Type Checked}}
\label{sec:motivation-landin-typed}
\begin{figure}[tb]
\begin{mdframed}
\centering
\resizebox{\textwidth}{!}{
\begin{tikzpicture}

\node[circular locs, ellipse, font=\large] (n1) at ([xshift=-5cm]current page.center) {\texttt{c:$\mu$z.Ref[(Unit  => Unit)$^z$]$^c$}};
\node[tracked lambda, aspect=4.5, cloud puffs=25, font=\large] (n2) at ([xshift=5cm]current page.center) {\texttt{f = $\lambda$x.(c...):(Unit => Unit)$^c$}};

\draw[nodearrow, bend left=15] (n1) to node[midway, below, yshift=-.2cm, font=\large] {\texttt{c := f}} (n2);
\draw[nodearrow, bend left=24] (n2) to node[midway, above, yshift=.3cm, font=\large] {\texttt{f : (Unit => Unit)$^\texttt{c}$}} (n1);

\end{tikzpicture}
}
\caption{Reachability graph of a well-typed \colorbox{purple!10}{cyclic
reference} assigned with a \colorbox{blue!10}{self-capturing function},
expressing a simple cycle through the store (this work).}
\label{fig:graph-cycle}
\end{mdframed}
\vspace{-3ex}
\end{figure}

Cyclic references enable our system to type check the Landin's knot encoding in
\Cref{fig:example-landin-circle}. \Cref{fig:graph-cycle} shows the reachability
graph of the well-typed cyclic reference @c@, which is updated with function @f@
that catures @c@. The update, \ie, @c := f@, is type checked by
\Cref{typing:t-sassgn-v} (in \Cref{fig:maybe-typing}).

When an assignment expression appears within an abstraction, where the argument
is a cyclic reference, we need to make sure that the actual argument being
applied to the abstraction also has a singleton variable, so that the expression
after $\beta$-reduction is still well-typed (see \Cref{typing:t-app} and
\Cref{typing:t-app-fresh} in \Cref{fig:maybe-typing}).\footnote{See
\Cref{typing:t-sassgn-l} and \Cref{typing:t-app-val} in
\Cref{fig:maybe-semantics} for the dynamic version of assignment and
application.}

We present the formal typing rules for cyclic references in \Cref{sec:formal},
and further discuss their implication and justify our design decisions in
\Cref{sec:limitation}.

\subsection{Shallow, Precise Reference Tracking}
\label{sec:motivation-shallow}

\Citet{DBLP:journals/pacmpl/WeiBJBR24} enforces a \textit{saturated, deep
dependency} on the reference qualifiers, where the reference's qualifier (outer)
always subsumes the referent's qualifier (inner). This design introduces rigid
coupling between the inner and the outer qualifier, leading to imprecise
reachability tracking.

Consider a reference @cell@ that contains an inner element @a@,
in \Citet{DBLP:journals/pacmpl/WeiBJBR24}'s system:

\begin{lstlisting}
  val a = ...           // : $\lstcm{\ty[a]{T}}$
  val cell = new Ref(a) // : $\lstcm{\ty[\texttt{cell,a}]{(\TRef[\ty[a]{T}])}}$   $\lsterrcm{\text{Imprecise! cell and a treated as shared}}$
\end{lstlisting}

Reference @cell@'s qualifier overapproximates its reachable sets by including
the qualifier of its enclosed element @a@. While such treatment is safe, it is
largely imprecise and precludes complex uses of references where a more granular
tracking precision is desired.

In \Citet{DBLP:journals/pacmpl/WeiBJBR24}'s system, one can define a function
(\eg, @newctx@) that attempts to circumvent outer qualifier propagation by
passing a reference as a fresh argument (via \Cref{typing:t-app-fresh}, see
\Cref{fig:maybe-typing}) and via partial application, passing another fresh
reference enclosing the same underlying value @inner@:

\begin{lstlisting}
  def newctx(c1' : $\ty[\QFresh]{\TRef[\ty[inner]{T}]}$)(c2' : $\ty[\QFresh]{\TRef[\ty[inner]{T}]}$) = ...  // attempt to treat c1' and c2' as separate
\end{lstlisting}

However, @newctx@ can never be invoked with arguments of their desired types
under \Citet{DBLP:journals/pacmpl/WeiBJBR24}'s system, as it demands that the
two separate references contain a shared value @inner@, violating the separation
constraint illustrated in \Cref{fig:graph-wei}. As a failed attempt, partially
applying @newctx@ to @c1@ results in a new function that captures @inner@, which
overlaps with the @c2@. Since @c2@ is required to be contextually fresh, it
cannot be applied as the second argument of @newctx@ (shown below to the left):

\begin{minipage}{0.5\linewidth}
\begin{lstlisting}
val inner = ...         // : $\lstcm{\ty[\texttt{inner}]{T}}$
val c1 = new Ref(inner) // : $\lstcm{\ty[\texttt{c1,inner}]{(\TRef[\ty[inner]{T}])}}$ 
val c2 = new Ref(inner) // : $\lstcm{\ty[\texttt{c2,inner}]{(\TRef[\ty[inner]{T}])}}$ 
newctx(c1)(c2) // $\lsterrcm{\text{Error! c1 overlaps c2}}$
\end{lstlisting}
\end{minipage}
\begin{minipage}{0.5\linewidth}
\begin{lstlisting}
val inner = ...         // : $\lstcm{\ty[\texttt{inner}]{T}}$
val c1 = new Ref(inner) // : $\lstcm{\ty[\texttt{c1}]{(\TRef[\ty[inner]{T}])}}$ 
val c2 = new Ref(inner) // : $\lstcm{\ty[\texttt{c2}]{(\TRef[\ty[inner]{T}])}}$ 
newctx(c1)(c2) // Okay, c1 and c2 are disjoint
\end{lstlisting}
\end{minipage}

To address this limitation, we disentangle the reference qualifier and the
referent qualifier, removing the inclusion requirement on the outer reference
qualifier. As a result, the same code is correctly type checked in \maybelang,
where the references @c1@ and @c2@ are disjoint, and the function @newctx@ can
be invoked with the desired arguments (shown above to the right).

We incorporate the shallow tracking mechanism with a novel notion of \rchty{} in
\maybelang, and present it formally in \Cref{sec:formal}, highlighting the
difference in reference introduction rule comparing to
\Citet{DBLP:journals/pacmpl/WeiBJBR24}'s system.  The shallow tracking mechanism
we propose scales well to nested references (\Cref{fig:graph-this}), where the
same invariants guaranteed by the deep tracking system remain valid
(\Cref{sec:formal-meta}).

\subsection{Reference Escaping via Controlled Imprecision}
\label{sec:motivation-escape}

Consider again the function \texttt{mkRef} in \Cref{fig:escape-old}, which
attempts to return a reference \texttt{c} that was created within the function  
body. This is not allowed in \citet{DBLP:journals/pacmpl/WeiBJBR24}'s system, 
because \texttt{c}'s referent qualifier contains \texttt{x}, a parameter local
to the function. In \maybelang, we introduce a controlled form of imprecision
(\Cref{fig:escape-new}). 

Instead of requiring that the returned reference \texttt{c} precisely track
\texttt{x}, we allow @x@ to \textit{escape} from the \rdcolor{read} path of its
referent qualifier to its outer qualifier. In other words, we replace \(x\) with
the binder \(z\), and simultaneously add @x@ to @c@'s outer qualifier. This way,
the system still tracks all reachable locations on its \rdcolor{read} path,
including the escaped qualifiers, which will be re-introduced upon dereferencing
(see \Cref{typing:t-sderef-dual} and \Cref{typing:t-esc} in
\Cref{fig:dual-component-semantics}). 

On the \wrcolor{write} path, we do not include the binder \(z\) to prevent
further writes to the escaped reference. Removal of \(x\) from the
\wrcolor{write} path is safe due to contravariance (see \Cref{sec:dual}). When
the write path is empty, we abbreviate @c@'s read only cyclic reference type
\(\mty[z]{x}{\SRf{T}{\qbot}{z}}\) as \(\mty[z]{x}{\TRef[T^\rdcolor{\&z}]}\) (see
\Cref{fig:dual-component-semantics}).

\paragraph{\textbf{Read-only references via reference widening}} Building on the
idea of read-only references, our system also supports read-only references
through reference widening, which restricts write access while still allowing
reads. For example, we can define a function that accepts a read-only reference
\texttt{r} by restricting its write qualifier to \(\qbot\), ensuring that no
tracked values can be written into it, while allowing reads:  

\begin{lstlisting}
  def useReadOnlyRef(r: $\mty[z]{r}{\SRf{\TRef[Int]}{\qbot}{z}}$) =
    println(!r)         // Okay, immutable access allowed
    val a = new Ref(42) // : $\lstcm{\ty[a]{\TRef[Int]}}$
    // !r := a          // $\lsterrcm{\text{Error! Write qualifier is empty}}$
\end{lstlisting}  

Notably, this read-only restriction applies only at the abstraction boundary,
where the reference is treated as read-only upon function application through
the function's domain type. The reference itself remains mutable and may be
freely updated outside the function:

\begin{lstlisting}
  val a = new Ref(10)     // : $\lstcm{\ty[a]{\TRef[Int]}}$
  val ref = new Ref(a)    // : $\lstcm{\ty[ref]{\TRef[\ty[a]{\TRef[Int]}]}}$
  useReadOnlyRef(ref)     // Okay, reference is safely upcasted before being passed to the function
  // $\lstcm{\ty[r]{\TRef[\ty[a]{\TRef[Int]}]} <: \mty[z]{r}{\SRf{\TRef[Int]}{\qbot}{z}}}$
  ref := ...              // Okay, ref is mutable outside the function
\end{lstlisting}

We formalize the dual-component reference types in \Cref{sec:dual} and provide
examples, including read-only references for untracked values, nested
escaping, and additional use cases, in \Cref{sec:casestudy-escape}. %
\section{\maybelang: Reachability Types with Cyclic Reference Type}\label{sec:formal}

\begin{figure}[t]\footnotesize
\begin{mdframed}
\judgement{Syntax}{\BOX{\maybelang}}\small\vspace{-10pt}
\[\begin{array}{l@{\qquad}l@{\qquad}l@{\qquad}l}
    x,y,z   & \in & \Var                                                                            & \text{Variables}           \\
    f,g,h   & \in & \Var                                                                            & \text{Function Variables}  \\
    X       & \in & \Var                                                                            & \text{Type Variables}      \\ [2ex]

    S,T,U,V & ::= & \TUnit \mid f(x: \ty{Q}) \to \ty{R} \mid \ext{\mr{}{Q}{}}                       &                            \\
            &     & \mid \TTop \mid X \mid \forall f(\ty[x]{X} <: Q). Q                             & \text{Types}               \\
    B       & ::= & \TUnit                                                                          & \text{Base Types}          \\
    t       & ::= & c \mid x \mid \lambda f(x).t \mid t~t \mid \tref~t \mid\ !~t \mid t \coloneqq t &                            \\
            &     & \mid \TLam{X}{x}{T}{q}{t} \mid \TApp{t}{Q}{}                                    & \text{Terms}               \\ [2ex]

    p,q,r,w & \in & \mathcal{P}_{\mathsf{fin}}(\Var \uplus \{ \QFresh \})                           & \text{Type Qualifiers}     \\
    O,P,Q,R & ::= & \ty[q]{T}                                                                       & \text{Qualified Types}     \\ [2ex]

    \flt    & \in & \mathcal{P}_{\mathsf{fin}}(\Var)                                                & \text{Observations}        \\
    \Gamma  & ::= & \varnothing\mid \Gamma, x : Q   \mid \Gamma, \ty[x]{X} <: Q                     & \text{Typing Environments} \\
\end{array}\]

\footnotesize
\judgement{Qualifier Notations}
\[\begin{array}{l@{\qquad}l@{\qquad}l@{\qquad}l}
    $\TRef~[Q]$       & := & $\mty{}{\TRef~[Q]}$ \qquad \text{if } x \notin \FV(Q) & \text{Reference Notation}             \\
    p,q               & := & p \qlub q                                              & \text{Qualifier Union}               \\
    q \ominus x       & := & q \setminus \{x\}                                      & \text{Qualifier Exclusion}           \\
    x                 & := & \{x\}                                                  & \text{Single Variable Qualifier}     \\
    \QFresh           & := & \{\QFresh\}                                            & \text{Single Fresh Qualifier}        \\
    \ext{\{\Var\}}    & := & \{\{x\} \mid x \in \Var\}                              & \text{Singleton Variable Qualifiers}    \\
    \ext{\Singleton}  & := & \{\Var\}                                               & \text{Singleton Qualifiers}            \\
\end{array}\]

\caption{The syntax of \maybelang. Additions to
\Citet{DBLP:journals/pacmpl/WeiBJBR24} are highlighted in \ext{\text{gray
boxes}}.}\label{fig:maybe-syntax}
\end{mdframed}
\vspace{-3ex}
\end{figure}

In this section, we formally present \maybelang that extends
\citet{DBLP:journals/pacmpl/WeiBJBR24}'s $\polylang{}$ calculus with features
discussed previously.  Differences from their system are highlighted in
\ext{\text{gray boxes}}.  We also use \oldcolor{gray} to indicate typing rules
in their system that are superseded by our new rules, aligning them side-by-side
for easy comparison.

\subsection{Syntax}
\label{sec:formal-syntax}

We present the formal syntax of \maybelang in \Cref{fig:maybe-syntax}. We
introduce our reference type that is of the form of $\mr[x]{p}{T}{q}$ , where
$\m x$ binds $x$ to the referent's qualifier $q$.  If $x \in q$, it allows the
referent to refer to its reference; otherwise, it denotes a \textit{non-cyclic
reference}, and can be abbreviated as $\ty[p]{\TRef~[\ty[q]{T}]}$, which is the
notation adopted  from \citet{DBLP:journals/pacmpl/WeiBJBR24}'s work, serving as
syntactic sugar in our system. Note that if $x \in q$, the full $\m x$ notation
must be used to avoid ambiguity. We refer readers to \Cref{appendix:syntax} for
the full syntax of \maybelang.

It is worth noting that although our $\m x$ notation coincides with other
systems, such as recursive types \citep{DBLP:books/daglib/0005958,
DBLP:conf/lics/AbadiF96} and DOT
\citep{DBLP:conf/oopsla/RompfA16,DBLP:conf/birthday/AminGORS16}, it has a
distinct meaning. See \Cref{sec:related} for the detailed elaboration.

\subsection{Static Typing}
\label{sec:formal-static}
\begin{figure}[t]\footnotesize
\begin{mdframed}
  \judgement{Term Typing}{\BOX{\strut\G[\flt] \ts t : \ty{Q}}}\\[1ex]
  \begin{minipage}[t]{.5\linewidth}\small\vspace{0pt}
    \typerule{t-var}{
      y : \ty[q]{T} \in \G\quad\quad y \in \flt
    }{
      \G[\flt] \ts y : \ty[y]{T}
    }
  \end{minipage}
  \begin{minipage}[t]{.5\linewidth}\small\vspace{0pt}
      \typerule{t-cst}{
        c \in B
      }{
        \G[\flt] \ts c : \ty[\qbot]{B}
      }
  \end{minipage}
  \vgap
  \begin{minipage}[t]{0.3\linewidth}\small\vspace{0pt}
    \oldcolor{
      \typerule{t-ref}{
          \G[\flt]\ts t : \ty[q]{T}\qquad \QFresh\notin q
        }{
          \G[\flt]\ts \tref~t : \ty[{q,\QFresh}]{(\TRef~\ty[q]{T})}
        }
    }
  \end{minipage}
  \begin{minipage}[t]{0.3\linewidth}\small\vspace{0pt}
      \typerule{t-sref}{
        \G[\flt]\ts t : \ty[q]{T}\qquad \QFresh\notin q
      }{
        \G[\flt]\ts \tref~t : \ty[\ext{\QFresh}]{(\TRef~\ty[q]{T})}
      }
  \end{minipage}
  \begin{minipage}[t]{0.3\linewidth}\small\vspace{0pt}
      \typerule{t-sref-2}{
        \G[\flt]\ts t : \ty[q]{T}\qquad \QFresh\notin q
      }{
        \G[\flt]\ts \tref~t : \ext{\mr{\QFresh}{T}{x,q}}
      }
  \end{minipage}
  \vgap
  \begin{minipage}[t]{.35\linewidth}\small\vspace{0pt}
    \oldcolor{
      \typerule{t-deref}{
        \G[\flt]\ts t : \ty[p]{(\TRef~\ty[q]{T})} \quad q\subq\flt
      }{
        \G[\flt]\ts !t : \ty[q]{T}
      }
    }
  \end{minipage}
  \begin{minipage}[t]{.6\linewidth}\small\vspace{0pt}
      \typerule{t-sderef}{
        \G[\flt]\ts t : \mr{p}{T}{q}
        \quad \ext{q\subq\flt,x} \quad \ext{x \notin\FV(T)}
      }{
        \G[\flt]\ts !t : \ty[{\ext{q[p/x]}}]{T}
      }
  \end{minipage}
  \vgap
  \begin{minipage}[t]{.27\linewidth}\small\vspace{0pt}
    \oldcolor{
      \typerule{t-assgn}{
        \G[\flt]\ts t_1 : \ty[p]{(\TRef~\ty[q]{T})} \\
        \G[\flt]\ts t_2 : \ty[q]{T}\quad\QFresh\notin q
      }{
        \G[\flt]\ts t_1 \coloneqq t_2 : \ty[\qbot]{\TUnit}
      }
    }
  \end{minipage}
  \begin{minipage}[t]{.26\linewidth}\small\vspace{0pt}
      \typerule{t-sassgn}{
        \G[\flt]\ts t_1 : \mr{p}{T}{q} \\
        \G[\flt]\ts t_2 : \ty[\ext{q \ominus x}]{T}
      }{
        \G[\flt]\ts t_1 \coloneqq t_2 : \ty[\qbot]{\TUnit}
      }
  \end{minipage}
  \begin{minipage}[t]{.35\linewidth}\small\vspace{0pt}
        \typerule{t-sassgn-v}{
          \G[\flt]\ts \ext{y} : \mr{p}{T}{\ext{q,x}} \\
          \G[\flt]\ts t_2 : \ty[\ext{y, q}]{T}
        }{
          \G[\flt]\ts \ext{y} \coloneqq t_2 : \ty[\qbot]{\TUnit}
        }
  \end{minipage}
  \vgap
  \begin{minipage}[t]{.45\linewidth}\small\vspace{0pt}
      \typerule{t-app}{
          \G[\flt]\ts t_1 : \ty[q]{\left(f(x: \ty[p]{T}) \to \ty{\ty[r]{U}}\right)} \\
          \G[\flt]\ts t_2 : \ty[p]{T}\qquad \QFresh\notin p \\ 
          r\subq\QFresh,\varphi,x,f \quad \ext{p \notin \Singleton \uplus \{\qbot\} \Rightarrow x\notin\FV(U)}
      }{
          \G[\flt]\ts t_1~t_2 : \ty{\ty[r]{U}}[p/x, q/f]
      }
  \end{minipage}%
  \begin{minipage}[t]{.55\linewidth}\small\vspace{0pt}
      \typerule{t-app-fresh}{
        \G[\flt]\ts t_1 : \ty[q]{\left(f(x: \ty[p\,{\overlap}\, q]{T}) \to \ty[r]{U}\right)}\quad
        \G[\flt]\ts t_2 : \ty[p]{T}\\ 
        \QFresh \in p \Rightarrow x\notin\FV(U)\quad \QFresh \in q \Rightarrow f\notin\FV(U) \\
        r\subq\QFresh,\varphi,x,f \quad \ext{{p \notin \Singleton \uplus \{\qbot\} \Rightarrow x\notin\FV(U)}}
      }{
        \G[\flt]\ts t_1~t_2 : (\ty[r]{U})[p/x, q/f]
      }
  \end{minipage}
  \vgap
  \begin{minipage}[t]{.35\linewidth}\small\vspace{0pt}
      \typerule{t-abs}{
        \cx[q,x,f]{(\G\ ,\ f: \ty{F}\ ,\ x: \ty{P})} \ts t : \ty{Q}\\
        \ty{F} = \ty[q]{\left(f(x: \ty{P}) \to \ty{Q}\right)}\quad q\subq \flt
      }{
        \G[\flt] \ts \lambda f(x).t : \ty{F}
      }
  \end{minipage}%
  \begin{minipage}[t]{.35\linewidth}\small\vspace{0pt}
      \typerule{t-tabs}{
        \cx[q,x,f]{\left(\G\ ,\ f: \ty{F}\ ,\ \ty[x]{X} <: \ty{P}\right)} \ts t : \ty{Q} \\
        F = \ty[q]{\left(\TAll{X}{x}{P}{}{Q}{}\right)} \quad q \subq \varphi
      }{
        \G[\flt] \ts \TLam{X}{x}{T_1}{q_1}{t} : F
      }
  \end{minipage}%
  \begin{minipage}[t]{.3\linewidth}\small\vspace{0pt}
      \typerule{t-sub}{
        \G[\flt]\ts t : \ty{Q} \qquad  \G\ts\ty{Q} <: \ty[q]{T}\\ 
        q\subq\flt,\QFresh
      }{
        \G[\flt]\ts t : \ty[q]{T}
      }
  \end{minipage}
  \vgap
  \begin{minipage}[t]{.45\linewidth}\small\vspace{0pt}
      \typerule{t-tapp}{
        \G[\flt] \ts t : \ty[q]{\left(\TAll{X}{x}{T}{p}{Q}{}\right)} \\
        \QFresh \notin p \quad  f\notin\FV(U) \\
        p \subseteq \varphi \quad r\subq\QFresh,\flt,x,f \quad Q = \ty[r]{U}
      }{
        \G[\flt] \ts t [ \ty[p]{T} ] : Q[\ty[p]T/\ty[x]{X}, q/f]
      }
  \end{minipage}%
  \begin{minipage}[t]{.55\linewidth}\small\vspace{0pt}
    \typerule{t-tapp-fresh}{
      \G[\flt] \ts t : \ty[q]{\left(\TAll{X}{x}{T}{p \overlap q}{Q}{}\right)} \\
      \QFresh \in p \Rightarrow x\notin\FV(U) \quad f\notin\FV(U) \\
      p \subseteq \varphi \quad r\subq\QFresh,\flt,x,f \quad Q = \ty[r]{U}
    }{
      \G[\flt] \ts t [ \ty[p]{T} ] : Q[\ty[p]{T}/\ty[x]{X}, q/f]
    }
  \end{minipage}
\caption{Typing rules of \maybelang. Differences from
\citet{DBLP:journals/pacmpl/WeiBJBR24} are highlighted in \ext{\text{gray
boxes}}. The \oldcolor{gray} rules are directly taken from
\citet{DBLP:journals/pacmpl/WeiBJBR24} for comparison purposes, and are not a
part of \maybelang. Note the differences in dereference, assignment, application
rules (\Cref{sec:motivation-cyclic}) and reference introduction rule
(\Cref{sec:motivation-shallow}). The qualifier overlap operator (\protect\overlap) used in
\Cref{typing:t-app-fresh} and \Cref{typing:t-tapp-fresh} is described in
\Cref{fig:maybe-qtrans}.}\label{fig:maybe-typing}

\end{mdframed}
\end{figure}
 
As in \citet{DBLP:journals/pacmpl/WeiBJBR24}'s work, the term typing judgment is written as: $\G[\flt] \ts t: Q$, as shown in \Cref{fig:maybe-typing}.
It means that term $t$ has type $Q$ and is allowed to access what are observable from $\flt$ in typing environment $\G$.
A type $Q$ is qualified with reachability qualifiers, and has the form $\ty[q]{T}$.

Following \citet{DBLP:journals/pacmpl/WeiBJBR24}, term typing assigns minimal qualifiers.
\Cref{typing:t-var} and \Cref{typing:t-cst} follow their systems.
In this section, we assume $\TUnit$ is the only base type in the system.

\subsubsection{Reference Introduction}\label{sec:typing-intro}

\Cref{typing:t-sref} and \Cref{typing:t-sref-2} introduce the reference type for
\textit{non-cyclic} and \textit{cyclic} references respectively.  Comparing with
\Cref{typing:t-ref}  in \citet{DBLP:journals/pacmpl/WeiBJBR24}'s system, 
\Cref{typing:t-sref} removes the deep dependency between the inner and the outer
qualifiers, by not propagating the referent qualifier (\ie, $q$) to its outer
qualifier.  Effectively, reference qualifiers in our system now denote \rchty{}
rather than transitive reachability, weakening the notion of reference
separation and enabling more precise reference tracking.  Through the revised
reference introduction rules, we maintain a shallow dependency between reference
and its inner referent, making reference outer qualifier track only its \rch{}
resources by default.

Similar to \citet{DBLP:journals/pacmpl/WeiBJBR24}, we disallow creating
references to fresh values to ensure sound reachability tracking.
Without this restriction, dereferencing such a reference multiple times would produce values
that, from the qualifier's perspective, appear to be distinct fresh resources
while actually referring to the same underlying resource.

\subsubsection{Dereference Rules}\label{sec:typing-deref}

\Cref{typing:t-sderef} is for dereferencing references, ensuring that the
retrieved value's qualifier propagates properly.  Importantly, if the referent
qualifier includes the cyclic reference type binder, \eg, $x$, then we
substitute $x$ with its outer qualifier $p$ in the resulting qualifier after
dereferencing. This substitution is necessary, as here $x$ acts as an
abstraction for the outer qualifier within the reference type. Once
dereferencing eliminates the reference type, this abstraction must be replaced
with the actual outer qualifier, as the resulting value no longer contains a
reference and thus cannot retain the abstraction.

Through dereferencing, we concretize this abstraction, making the
outer reference qualifier explicit in the resulting type. However, such an
abstraction is limited in that the bound variables can refer only to their
immediate outer qualifier, rather than qualifiers at arbitrary depth, as seen in
lambda abstractions (discussed in \Cref{sec:limitation}). 

For non-cyclic reference types, where the bound variable @x@ is absent from the
referent qualifier, this rule generalizes rule \Cref{typing:t-deref} in
\citet{DBLP:journals/pacmpl/WeiBJBR24}, simply retrieving the referent
qualifier.

\subsubsection{Assignment Rules}\label{sec:typing-assgn}

\Cref{typing:t-sassgn} and \Cref{typing:t-sassgn-v} are for safe assignment of
values to references.  \Cref{typing:t-sassgn} exhibits the same semantics as
\Cref{typing:t-assgn} in \Citet{DBLP:journals/pacmpl/WeiBJBR24}'s system when
assignee is a non-cyclic reference. When the assignee is a cyclic reference, the
cyclic binder is simply discarded, and a non-cyclic assignment is performed.

\Cref{typing:t-sassgn-v} is used for cyclic reference assignments, allowing
values to reach the reference's outer qualifier.  The system permits this
assignment only when the assignee is a variable and the assigned term has a
singleton qualifier matching the assignee's name. Allowing assignment without
enforcing a variable-form assignee and a matching singleton qualifier, as in
\Cref{typing:t-sassgn-err} (see \Cref{sec:motivation-cyclic}), would enable
assignments with arbitrary qualifiers, violating the intended type discipline.
This is because \Cref{typing:q-sub} allows arbitrary upcasting of a reference's
outer qualifier, and including a rule like \Cref{typing:t-sassgn-err} would lead
to inconsistencies in qualifier tracking.

Why, then, is it safe to restrict assignment to term forms that are variables?
By \Cref{typing:t-var}, a variable is always assigned a singleton qualifier that
matches its name. Although this variable's qualifier can be upcast
in an arbitrary way, the singleton qualifier is its most precise tracking prior to any
upcasting. Thus, restricting the assigned term's qualifier to a matching
singleton ensures that it refers to the intended reference and cannot be widened
to include other resources. 

In summary, for cyclic reference assignment, we determine the assigned term's
qualifier based on the assignee's term form, rather than its qualifier. Unlike
qualifiers, which can be upcast, term forms remain fixed, ensuring correctness
in assignment. Thus, \Cref{typing:t-sassgn-v} achieves a balance between
correctness and precision while successfully supporting cyclic reference
assignment.

\subsubsection{Application Rules}\label{sec:typing-app}

Similar to \citet{DBLP:journals/pacmpl/WeiBJBR24}, \maybelang defines two
variants of the application rule: \Cref{typing:t-app} for precise applications
(\ie, ``non-fresh'') and \Cref{typing:t-app-fresh} for growing applications (\ie,
``fresh'').

\Cref{typing:t-app} applies to non-fresh applications, where the argument's
qualifier is fully observable and explicitly specified. In this case, the
function's return type can depend on the function argument qualifier, and the
application performs a deep, precise substitution on the argument qualifier.  To
ensure correct deep substitution of qualifiers in the presence of cyclic
references, we impose an additional constraint: the argument qualifier must not
appear in the return type, unless it belongs to a restricted set of qualifiers
-- empty or singleton qualifiers (denoted by $\Singleton \uplus \{\qbot\}$) --
with which deep substitution is safe.

If the argument does not appear in the return type, type preservation is
trivially maintained since no type-level substitution is required.  If the
argument qualifier \textit{does} appear in the return type, it must be an empty
or singleton qualifier to ensure valid deep substitution. 

Why is this constraint necessary? Suppose a cyclic reference with non-empty,
non-singleton qualifier is applied to an abstraction, we need to ``shrink'' the
argument qualifier into a singleton qualifier to accommodate potential cyclic
assignment to the function's argument in the abstraction body (see
\Cref{lem:subst_term}). And if the argument qualifier also occurs in the
function's return type, we need to ``grow'' back the argument qualifier in the
return type to its original form to ensure type preservation (see
\Cref{sec:formal-meta}). But this is not always possible, as the argument
qualifier could occur in contravariant positions of the return type.

Admittedly, enforcing the non-occurrence constraint of argument qualifier in
function's return type for deep-substitution limits expressiveness, but this is
necessary to preserve type soundness. We further justify this limitation in
\Cref{sec:limitation-new}.

Deep substitution remains valid under the following cases, where the argument
qualifier is: (1) an empty qualifier ($p\in \{\qbot\}$), or a singleton
qualifier ($p \in \Singleton$).

Thus, when the argument qualifier is an empty or singleton qualifier, deep
substitution in \Cref{typing:t-app} remains valid because the qualifier is small
enough to be directly substituted using the substitution lemma (see
\Cref{lem:subst_term}).

Rule \Cref{typing:t-app-fresh} extends the constraints of \Cref{typing:t-app} to
fresh applications, enforcing the same non-occurrence constraint when the
qualifier is not an empty or singleton qualifier. Additionally, a similar
restriction applies to applications involving store locations (see
\Cref{sec:formal-dynamic}).

\subsubsection{Type Polymorphism}\label{sec:typing-poly}

Rules \Cref{typing:t-tapp} and \Cref{typing:t-tapp-fresh} extend the type
application rule to support type and qualifier polymorphism, similar to
\Citet{DBLP:journals/pacmpl/WeiBJBR24}, where a type abstraction can be typed
with \Cref{typing:t-tabs}. Unlike function applications, type applications
impose no restrictions on the argument qualifier, as types do not appear on the
left-hand side of assignment operations, eliminating concerns about improper
substitution.

\begin{figure}[t]\footnotesize
\begin{mdframed}
\judgement{Subtyping}{\BOX{\strut\G \ts q <: q}\ \BOX{\strut \G\ts\ty{T} <: \ty{T}}\ \BOX{\strut \G\ts\ty{Q} <: \ty{Q}}}\\[1ex]
\begin{minipage}[t]{.37\linewidth}\small\vspace{0pt}
 \typerule{s-base}{\ \\}{
    \G\ts\ty{B} <: \ty{B}
  }
  \vgap
  \typerule{s-trans}{
    \G\ts\ty{T} <: \ty{S} \quad
    \G\ts\ty{S} <: \ty{U}
  }{
    \G\ts\ty{T} <: \ty{U}
  }
\end{minipage}%
\begin{minipage}[t]{.03\linewidth}
\hspace{1pt}
\end{minipage}%
\begin{minipage}[t]{.6\linewidth}\small\vspace{0pt}
  \typerule{s-ref}{
    \G\ts\ty{S} <: \ty{T}  \quad
    \G\ts\ty{T} <: \ty{S}\quad q\subq\DOM(\Gamma)
  }{
    \G\ts\ty{\mr[z]{q}{S}{p}} <: \ty{\mr[z]{q}{T}{p}}
  }
  \vgap
  \typerule{s-fun}{
    \G\ts\ty{P} <: \ty{O} \\
     \G\, ,\, f : \ty[\QFresh]{(f(x : O)\to Q)}\, ,\, x : \ty{P}\ts \ty{Q} <: \ty{R}
  }{
    \G\ts\ty{f(x: \ty{O}) \to \ty{Q}} <: \ty{f(x: \ty{P}) \to \ty{R}}
  }
\end{minipage}
\begin{minipage}[t]{0.3\linewidth}\vspace{0pt}
  \typerule{q-sub}{
    p\subq q\subq \QFresh,\dom(\G)
  }{
    \G\ts p <: q
  }
\vgap
  \typerule{q-cong}{\G\ts q_1 <: q_2}{
    \G\ts p, q_1 <: p, q_2
  }
\end{minipage}
\begin{minipage}[t]{0.3\linewidth}\vspace{0pt}
  \typerule{q-self}{
    f : \ty[q]{T}\in\G \quad \QFresh\notin q
  }{
    \G\ts q,f <: f
  }
  \vgap
  \typerule{q-var}{
    x : \ty[q]{T}\in\G \quad \QFresh\notin q
  }{
    \G\ts x <: q
  }
\end{minipage}
\begin{minipage}[t]{0.4\linewidth}\vspace{0pt}
    \typerule{q-qvar}{
      \ty[x]{X} <: \ty[q]{T} \in \Gamma \quad
      \QFresh \notin q
    }{
      \Gamma \ts p,x <: p,q
    }
    \vgap
    \typerule{q-trans}{
      \G\ts p <: q \quad
      \G\ts q <: r
    }{
      \G\ts p <: r
    }
\end{minipage}
\begin{minipage}[t]{0.35\linewidth}\vspace{0pt}
    \typerule{s-tvar}{
      \ty[x]{X} <: \ty[q]{T} \in \Gamma
    }{
      \Gamma \ts X <: T
    }
\end{minipage}
\begin{minipage}[t]{0.2\linewidth}\vspace{0pt}
    \typerule{s-top}{
      \
    }{
      \Gamma \ts T <: \TTop
    }
\end{minipage}
\begin{minipage}[t]{0.4\linewidth}\vspace{0pt}
  \typerule{sq-sub}{
    \G\ts\ty{S} <: \ty{T}\quad\quad \G\ts p <: q
  }{
    \G\ts\ty[p]{S} <: \ty[q]{T}
  }
\end{minipage}
\caption{Subtyping rules of \maybelang.}\label{fig:maybe-subtyping}
\end{mdframed}
\vspace{-3ex}
\end{figure}

 \subsubsection{Subtyping}\label{sec:typing-sub}

\Cref{fig:maybe-subtyping} presents the subtyping rules for \maybelang. Similar
to \citet{DBLP:journals/pacmpl/WeiBJBR24}, our subtyping system distinguishes
between qualifier subtyping, ordinary type subtyping, and qualified type
subtyping. 

Qualifier subtyping uses the standard subset inclusion
(\Cref{typing:q-sub}) along with two contextual reasoning rules
\Cref{typing:q-self} and \Cref{typing:q-var}. Rule \Cref{typing:q-qvar}
introduces subtyping to qualifier variables, while \Cref{typing:s-tvar} enables
type variable subtyping. \Cref{typing:q-cong} and \Cref{typing:q-trans} ensures
congruence and transitivity for qualifier subtyping.

Ordinary type subtyping includes invariance for references (\Cref{typing:s-ref})
and standard contravariant/covariant subtyping for functions
(\Cref{typing:s-fun}), with careful scoping of the function self-reference.

Finally, qualified type subtyping (\Cref{typing:sq-sub}) decomposes into
subtyping over the underlying types and their qualifiers. All subtyping rules
are unchanged from prior work and are fully mechanized. We refer readers to
\Citet{DBLP:journals/pacmpl/WeiBJBR24,DBLP:journals/pacmpl/BaoWBJHR21} for the
detailed explanation of these rules.

\subsection{Dynamic Typing} \label{sec:formal-dynamic}
\begin{figure}\footnotesize
\begin{mdframed}
\judgement{Term Typing (with Store Typing)}{\BOX{\cx[\flt]{[\Gamma\mid\Sigma]} \ts t : \ty{Q}}}
\[\begin{array}{l@{\qquad}l@{\qquad}l@{\qquad}l}
	\ell       & \in & \Loc                                                                        & \text{Locations}                     \\
	t          & ::= & \cdots \mid \ell                                                            & \text{Terms}                         \\
	v          & ::= & \lambda f(x).t \mid {c} \mid {\ell} \mid \tunit \mid \Lambda f(\ty[x]{X}).t & \text{Values}                        \\ [2ex]
	p,q,r      & \in & \mathcal{P}_{\mathsf{fin}}(\Var \uplus \Loc \uplus \{ \QFresh \})           & \text{Qualifiers}                    \\
	\flt       & \in & \mathcal{P}_{\mathsf{fin}}(\Var \uplus \Loc)                                & \text{Observations}                  \\
	\Sigma     & ::= & \varnothing \mid \Sigma,\ell : \ext{\mty[x]{q,x}{T}} & \text{Store Typing}                  \\ [2ex]

	\ty[q]{T}  & :=  & \mty[x]{q}{T} \qquad \text{if } x \notin q                                  & \text{Store Typing Notation} 			      \\
	\{\Loc\}   & :=  & \{\{\ell\} \mid \ell \in \Loc\}                                             & \text{Singleton Location Qualifiers} \\
	\Singleton & :=  & \{\Var\} \uplus \{\Loc\}                                                    & \text{Singleton Qualifiers}          \\
\end{array}
\]

\begin{minipage}[t]{.45\linewidth}\small\vspace{0pt}
\typerule{t-loc}{
    \Sigma(\ell) = \ty[q]{T}\qquad q\subq\DOM(\Sigma) \\
	\FV(T)=\varnothing\quad\FTV(T)=\varnothing \quad q,\ell \subq \flt
  }{
    \cx[\flt]{[\G\mid\Sigma]}\ts \ell : \ty[\ell]{\TRef~[\ty[q]{T}]}
  }
\end{minipage}
\begin{minipage}[t]{.03\linewidth}\end{minipage}
\begin{minipage}[t]{.51\linewidth}\small\vspace{0pt}
  \typerule{t-sassgn-l}
  {
    \cx[\flt]{[\G\mid\Sigma]}\ts \ell : \mr{p,\ell}{T}{q} \\
    \cx[\flt]{[\G\mid\Sigma]}\ts t_2 : \ty[q,\ell]{T}\quad
    \QFresh\notin q \quad x \in q
  }{
    \cx[\flt]{[\G\mid\Sigma]}\ts \ell \coloneqq t_2 : \ty[\qbot]{\TUnit}
  }
\end{minipage}
\vspace{5pt}

\begin{minipage}[t]{.53\linewidth}\small\vspace{0pt}
    \typerule{t-app-val}{
      \cx[\flt]{[\G\mid\Sigma]}\ts t_1 : \ty[q]{\left(f(x: \ty[p]{T}) \to \ty{Q}\right)} \quad \cx[\flt]{[\G\mid\Sigma]}\ts v : \ty[p]{T}\\
      Q = \ty[r]{U} \quad r\subq\QFresh,\varphi,x,f \quad \ext{p \notin \Singleton \uplus \{\qbot\} \Rightarrow v \neq \ell}
    }{
      \cx[\flt]{[\G\mid\Sigma]}\ts t_1~v : \ty{Q}[p/x, q/f]
    }
\end{minipage}
\begin{minipage}[t]{.03\linewidth}\end{minipage}
\begin{minipage}[t]{.40\linewidth}\small\vspace{0pt}
\typerule{t-sloc}{
    \Sigma(\ell) = \ext{\mty[x]{q,x}{T}}\quad q\subq\DOM(\Sigma) \\
	\FV(T)=\varnothing\quad\FTV(T)=\varnothing \quad q,\ell \subq \flt
  }{
    \cx[\flt]{[\G\mid\Sigma]}\ts \ell : \ext{\mr[x]{\ell}{T}{q,x}}
  }
\end{minipage}

\vgap

\judgement{Well-Formed and Well-Typed Stores}{\BOX{\Gamma \mid \Sigma \vdash \sigma}\ \BOX{\WF{\Sigma}}}
  $$\begin{array}{lll}
    \cx[\varphi]{[\Gamma \mid \Sigma]} \vdash \sigma & := &
      \varphi \subseteq \DOM(\sigma) \subseteq \DOM(\Sigma) \land
      \forall \ell \in \varphi, \cx[\varphi]{[\Gamma \mid \Sigma]} \vdash \sigma(\ell) : \Sigma(\ell) \\
    \Gamma \mid \Sigma \vdash \sigma & := &
      \cx[\DOM(\Sigma)]{[\Gamma \mid \Sigma]} \vdash \sigma
  \end{array}
  $$

  \begin{minipage}[t]{\linewidth}\small\vspace{0pt}
    \begin{minipage}[t]{.2\linewidth}\small\vspace{0pt}
    \typerule{st-emp}{\ }{\WF{\varnothing}}\qquad\qquad
    \end{minipage}
    \begin{minipage}[t]{.7\linewidth}\small\vspace{0pt}
    \typerule{st-con}
      {\WF{\Sigma}\quad \FV(T)=\varnothing \quad
      \FTV(T)=\varnothing \quad
      q \in \DOM(\Sigma) \quad
      \ell \notin\DOM(\Sigma) }{\WF{\Sigma\, ,\, \ell : \ty[q]{T}}}
    \end{minipage}
  \end{minipage}
  \begin{minipage}[t]{\linewidth}\small\vspace{0pt}
    \center
    \begin{minipage}[t]{.7\linewidth}\small\vspace{0pt}
    \typerule{st-scon}
      {\WF{\Sigma}\quad \FV(T)=\varnothing \quad
      \FTV(T)=\varnothing \quad
      q \in \DOM(\Sigma) \quad
      \ell \notin\DOM(\Sigma) }{\WF{\Sigma\, ,\, \ext{\ell : \mty[x]{q,x}{T}}}}
    \end{minipage}
  \end{minipage}

\caption{Extension with store typings for \maybelang. Store typing for cyclic
references are highlighted in \ext{\text{gray boxes}}.  In contrast to
\citet{DBLP:journals/pacmpl/WeiBJBR24}, we do not require saturated qualifiers
on well-formed stores. In the dynamic semantics, singleton qualifiers
($\Singleton$) are extended to include single locations apart from single
variables.
}\label{fig:maybe-semantics}
\end{mdframed}
\vspace{-3ex}
\end{figure}
 The \maybelang-calculus follows the standard call-by-value reduction for the
$\lambda$-calculus with mutable references.  
\Cref{fig:maybe-semantics} shows the dynamic typing rules, which incorporate
term typing and subtyping with store typing $\Sigma$.

The reduction rules are standard, and can be found in
Appendix~\Cref{appendix:semantics}.

\Cref{typing:t-sassgn-l} introduces a dynamic typing rule for cyclic
assignments, where the assignee is a single location. This rule is the dynamic
counterpart of \Cref{typing:t-sassgn-v}, where cyclic assignment expression
stays well-typed after the reference being assigned to is substituted into a
location.

\Cref{typing:t-app-val} extends typing to applications within abstraction bodies
at runtime, where an argument variable is substituted with a value. As in rule
\Cref{typing:t-app}, the empty or singleton qualifier constraint applies to
argument qualifiers. Singleton qualifiers now also extend to \textit{singleton
location qualifiers}.

When the argument qualifier in \Cref{typing:t-app-val} is \textit{not} a
singleton qualifier, the argument cannot be a single location, as demanded by
the substitution lemma (see \Cref{sec:formal-substitution}). The call-by-value
semantics allows us to relax the static typing constraint on the non-occurrence
of the argument qualifier in the function's return type (\ie, $x \not \in
\FV(U)$), as per \Cref{typing:t-app}. Instead, we impose a weaker requirement:
the argument must not be a location value (\ie, $v \neq \ell$). This is enough
because call-by-value semantics ensures that arguments are reduced to values,
and the argument can have a reference type \textit{only if} it is reduced to a
location value (\ie, $\ell$) at call site.

This restriction prevents improper substitution by ensuring locations always
have their corresponding singleton qualifier when applied to abstractions. In
contrast, static typing must impose the non-occurrence constraint because
arguments may include arbitrary terms beyond values, and their types may involve
type abstractions, which cannot be checked syntactically. 

Unlike \citet{DBLP:journals/pacmpl/WeiBJBR24}'s $\polylang{}$-calculus,
\maybelang drops the saturation constraint for qualifiers in the store,
extending store typing to support both cyclic references and non-cyclic
references. 

\subsection{Precise Transitive Closure Computation}\label{sec:formal-mechanization}
\begin{figure}[t]
\begin{mdframed}\footnotesize

\judgement{Variable Reachability and Lookup}\BOX{{\color{gray}\G\vdash}\, x \reaches y}\ \BOX{{\color{gray}\G\vdash}\, x \reaches^* y}\ \BOX{{\color{gray}\G\vdash}\, \qtrans{x}}%
\[\begin{array}{l@{\ \,}c@{\ \,}l@{\qquad\qquad\qquad\ \ }l}
  {\color{gray}\G\vdash}\, x \reaches y & \Leftrightarrow  & x : T^{q,y} \in \G & \text{Variable Reachability}  \\ [1.1ex]
  {\color{gray}\G\vdash}\, x \reaches^* y & \Leftrightarrow  & x \reaches y \lor \exists z, x \reaches z \land z \reaches^* y & \text{Variable Transitive Reachability}  \\ [1.1ex]
  {\color{gray}\G\vdash}\, \qtrans{x} & := & \left\{\, y \mid x \reaches^* y\, \right\} & \text{Variable Saturation} \\[1.1ex] 
\end{array}\]

\judgement{Qualifier Transitive Lookup}{\BOX{{\color{gray}\G\vdash}\, \qtrans[n]{q}}\ \BOX{{\color{gray}\G\vdash}\, \qtrans{q}}}
\[\begin{array}{l@{\ \,}c@{\ \,}l@{\qquad\qquad\qquad\ \ }l}
	{\color{gray}\G\vdash}\, \qtrans[0]{q}   & := & q                                                      & \text{Qualifier Transitive Lookup Base Case}  \\ [1.1ex]
	{\color{gray}\G\vdash}\, \qtrans[n+1]{q} & := & \qtrans[n]{((\bigcup_{x \in q}\, \left\{\, y \mid x \reaches y\, \right\}) \cup q)}  & \text{Qualifier Transitive Lookup Inductive Case} \\ [1.1ex]
	{\color{gray}\G\vdash}\, \qtrans{q}      & := & \qtrans[\norm{\G}]{q}                                  & \text{Qualifier Full Transitive Lookup}       \\ [1.1ex]
\end{array}\]

\begin{minipage}[t]{\linewidth}\vspace{0pt}
\judgement{Qualifier Notations}{\BOX{{\color{gray}\G\vdash}\, \oldcolor{\psat{q}}}\ \BOX{{\color{gray}\G\vdash}\, \dsat{q}}\ \BOX{{\color{gray}\G\vdash}\, p \overlap q}}
\[\begin{array}{l@{\ \,}c@{\ \,}llp{.5\textwidth}}
	{\color{gray}\G\vdash}\, \oldcolor{\psat{q}} & \oldcolor{:=} & \oldcolor{(\bigcup_{x \in q} \qtrans{x}) \subq q} & \oldcolor{\text{Propositional Qualifier Saturation}} \\ [1.1ex]
	{\color{gray}\G\vdash}\, {\dsat{q}}            & := & \qtrans{q} = q & \text{Deterministic Qualifier Saturation} 	\\ [1.1ex]
	{\color{gray}\G\vdash}\, p \overlap q & := & \QFresh, {(\qtrans{p} \qglb\, \qtrans{q})} & \text{Qualifier Overlap}\\ [1.1ex]
\end{array}\]
\end{minipage}

\caption{Qualifier Transitive Lookup.
  The \oldcolor{propositional qualifier saturation predicate (\(\psat{q}\))}
  in \Citet{DBLP:journals/pacmpl/WeiBJBR24} is superseded by the deterministic
  qualifier saturation predicate (\(\dsat{q}\)) in the mechanization of
  \maybelang-calculus, which is defined in terms of the deterministic transitive
  lookup computation (\(\qtrans{q}\)). 
  }\label{fig:maybe-qtrans}
\end{mdframed}
\vspace{-3ex}
\end{figure}

 In \Citet{DBLP:journals/pacmpl/WeiBJBR24}'s mechanization, the concept of
transitive closure is encoded indirectly via a ``saturation'' predicate, which
specifies that a qualifier will reach no more than itself through the
corresponding store. Coupled with universal quantification, this yields an
overapproximate notion of transitive closure, where a \emph{saturated superset}
of $p$ is used to characterize $p$'s transitive closure. While a precise
specification of transitive closure as the \emph{smallest saturated superset} is
possible, doing so requires second-order quantification and complicates
mechanization.

We replace this definition by replacing the transitive closure with a
\textit{deterministic} definition (\Cref{fig:maybe-qtrans}). By using a ``fuel''
measure ($n$) that is set to the length of the typing context ($\norm{\G}$), the
computation is guaranteed to terminate while fully capturing the transitive
closure of a qualifier, even when the typing context contains cycles.

We define standard monotonicity, commutativity, and distributivity properties of
the transitive closure operation with respect to other operations on sets and
qualifiers (\eg, splice, substitution, \etc). In our Rocq mechanization, we
further show that the computational definition subsumes the saturation by
proving an equivalence theorem between the two (see
\Cref{lem:qenv-saturated-iff}).

\newcommand{\LocType}[3]{#1 \vdash #2 : #3}

\newcommand{\Substq}[5]{%
  \ensuremath{[#1\mid#2] \vdash #4 \xRightarrow{#3} #5}%
}

\begin{figure}[t]
\begin{mdframed}\footnotesize

\judgement{Location Typing}\BOX{\LocType{\Sigma}{\ell}{\ty[d]{T}}}

\begin{minipage}[t]{0.2\linewidth}\vspace{0pt}
\typerule{vl-loc}{
  \
}{
  \LocType{\Sigma}{\ell}{\ty[\ell]{\SR{T}{q}}}
}
\end{minipage}%
\begin{minipage}[t]{0.25\linewidth}\vspace{0pt}
\typerule{vl-sloc}{
  \
}{
  \LocType{\Sigma}{\ell}{\mr[x]{\ell}{T}{q}}
}
\end{minipage}%
\begin{minipage}[t]{0.08\linewidth}\vspace{0pt}
\typerule{vl-top}{
  \
}{
  \LocType{\Sigma}{\ell}{\ty[d]{\TTop}}
}
\end{minipage}%
\begin{minipage}[t]{0.45\linewidth}\vspace{0pt}
\typerule{vl-store}{
  q \subq x, \DOM(\Sigma) \;\;
  \Sigma(\ell) = \ty[q]{T} \;\;
  p \subq q
}{
  \LocType{\Sigma}{\ell}{\mr[x]{\ell, {p[\qbot/x]}}{T}{q}}
}
\end{minipage}%

\judgement{Qualifier Substitution}\BOX{\Substq{\G}{\Sigma}{T}{q}{p}}

\begin{minipage}[t]{0.15\linewidth}\vspace{0pt}
\typerule{sn-exact}{
  \QFresh \not\in q
}{
  \Substq{\G}{\Sigma}{T}{q}{q}
}
\end{minipage}%
\begin{minipage}[t]{0.2\linewidth}\vspace{0pt}
\typerule{sn-grow}{
  \QFresh \not\in p \;\;
  q \subq \DOM(\G,\Sigma)
}{
  \Substq{\G}{\Sigma} {T} {q \overlap p} {p}
}
\end{minipage}%
\begin{minipage}[t]{0.27\linewidth}\vspace{0pt}
\typerule{sn-loc}{
  \LocType{\Sigma}{\ell}{\ty[p]{T}} \;\;
  \QFresh \not\in p \;\;
  p \subq \phi
}{
  \Substq{\G}{\Sigma}{T}{\phi}{p}
}
\end{minipage}%
\begin{minipage}[t]{0.36\linewidth}\vspace{0pt}
\typerule{sn-loc-grow}{
  \LocType{\Sigma}{\ell}{\ty[p]{T}} \;\;
  \QFresh \not\in \phi \;\;
  q,\phi \subq \DOM(\G,\Sigma) \;\;
  p \subq \phi
}{
  \Substq{\G}{\Sigma}{T}{q \overlap \phi}{p}
}
\end{minipage}%
\caption{Qualifier Substitution.}\label{fig:maybe-substq}
\end{mdframed}
\vspace{-3ex}
\end{figure}

 \subsection{Metatheory} 
\label{sec:formal-meta}

We list key lemmas and theorems that establish the soundness of \maybelang. The
proof structure follows that of \Citet{DBLP:journals/pacmpl/WeiBJBR24}.

\subsubsection{Saturation Lemma}
\label{sec:formal-saturation}
\begin{lemma}[Equivalence of Deterministic and Propositional Saturation]
\label{lem:qenv-saturated-iff}
For any environment $\Gamma$ and qualifier $q$, we have ${\color{gray}\G\vdash}\, \psat{q} \;\Longleftrightarrow\; {\color{gray}\G\vdash}\, \dsat{q}$

\end{lemma}

\subsubsection{Substitution Lemma}
\label{sec:formal-substitution}

We now describe qualifier substitution, a key relation that defines how
qualifiers can be substituted in $\beta$-reduction, which leads to the
substitution lemma. 

\Cref{fig:maybe-substq} lists the four cases where qualifier substitution can be
applied. \Cref{typing:sn-exact} is the case of \textit{precise substitution} where the
qualifier \(q\) is substituted for itself. This occurs for function parameters
in \Cref{typing:t-app} or for a function's self-reference \(f\) in
\Cref{typing:t-app} and \Cref{typing:t-app-fresh}. 
\Cref{typing:sn-grow} is the case of \textit{growing substitution} where the
argument qualifier \(p\) overlaps with the function qualifier \(q\), growing the
result by \(p \setminus ({\color{gray}\G\vdash}\, \qtrans{q})\). 
\Cref{typing:sn-loc} is the case of \textit{precise substitution with a location
qualifier}, where a single location qualifier can be substituted similar to
\Cref{typing:sn-exact}.
Lastly, \Cref{typing:sn-loc-grow} is the case of \textit{growing substitution
with a location qualifier}, where we first grow the location qualifier \(p\)
arbitrarily to some \(\phi\) and overlap it with the function the function
qualifier \(q\).  similar to \Cref{typing:sn-grow}.

\begin{lemma}[Substitution preserves transitive lookup]\label{lem:subst_preserves_trans}
\label{lem:q_trans_subst1_tenv_subq}
If $p \subq \QFresh, \dom(\Sigma)$, 
  and $\Substq{\G, x : \ty[q]{T}}{\Sigma}{T}{q}{p}$, 
  then under substitution \( \theta = [p/x] \), 
  $
    ({\color{gray}\G\theta\vdash}\, \qtrans{r\theta}) \subq ({\color{gray}\G, x : \ty[q]{T}\vdash}\, \qtrans{r})\theta.
  $
\end{lemma}

\begin{proof}
   We proceed by applying transitivity of subtyping, followed by commutativity
   of transitive lookup under substitution, monotonicity of qualifier
   substitution, and environment weakening properties. The proof concludes by
   applying subqualifier transitivity and narrowing arguments.
\end{proof}

For both qualifier substitution (see \Cref{fig:maybe-substq}) and term
substitution (\Cref{lem:subst_term}), we require the location to be well-typed
according to a location typing relation.

\Cref{typing:vl-loc} and \Cref{typing:vl-sloc} requires that the location
qualifier of a location value $\ell$ must be a singleton qualifier $\ell$, when
its type is a cyclic reference type or a non-cyclic reference type respectively.
\Cref{typing:vl-top} allows a location to have arbitrary qualifier when its type
is upcast to $\top$.  Lastly, \Cref{typing:vl-store} allows a location $\ell$ of
cyclic reference type to contain an additional qualifier $p$ that is reachable
from the store typing $\Sigma$ at location $\ell$. This last case is crucial for
substituting partially or fully escaped location values.

\begin{lemma}[Top-Level Term Substitution]\label{lem:subst_term}
  Suppose the following typing judgments hold:
    $\cx[\flt]{[x:\ty[q]{T},\G\mid\Sigma]} \ts t : \ty{Q}$,
    $\cx[p]{[\varnothing\mid\Sigma]} \ts v : \ty[p]{T}$.
  Additionally, assume:
    \(({\color{gray}x:\ty[q]{T},\G \vdash}\ p \overlap \flt) \subq q\), and
    \(p, q \subq \QFresh, \dom(\Sigma)\), and
    \(\Substq{\G}{\Sigma}{T}{q}{p}\), and
    \(\exists \ell, v = \ell \implies \LocType{\Sigma}{v}{p} \).
  Then, under substitution $\theta = [p/x]$, we have:
    \(\cx[\flt\theta]{[\Gamma\theta\mid\Sigma]} \ts t[v/x] : \ty{Q}\theta\).
\end{lemma}

\begin{proof}
  We proceed by induction on the derivation  
  \(\cx[\flt]{[x:\ty[q]{T},\G\mid\Sigma]} \ts t : \ty{Q}\).

  In the case of \Cref{typing:t-app-fresh}, the induction hypothesis requires  
  \Cref{lem:subst_preserves_trans} to establish  
  \((p \overlap q)\theta \subq p\theta \overlap q\theta\).

  In the case of \Cref{typing:t-sassgn-v}, we perform case analysis on the variable assignee.  
  If the variable is the one being substituted, we use the fact that the substituted value  
  is a reference, allowing us to conclude that it is a location.  
  We then apply \Cref{typing:t-sassgn-l} to complete the case.  
  If the variable is different from the one being substituted, the proof follows directly  
  from the induction hypothesis.
  Other cases follow straightforwardly from the induction hypothesis.
\end{proof}

\noindent

Since cyclic assignment (\Cref{typing:t-sassgn-v}) can occur inside an
abstraction body, if the assigned term's qualifier contains the function
argument qualifier, the latter must be substituted with an \textit{empty or
singleton qualifier} to ensure that the cyclic assignment remains well-typed
after qualifier substitution. 

As we consider only top-level, closed values in substitution, this constraint
becomes relevant only when the substituted term is a location value (\eg
$\ell$), as it could potentially serve as the assignee in a cyclic assignment
operation. This requirement is captured by the premise \((\exists \ell, v =
\ell \to p = \ell)\).

\subsubsection{Main Soundness Result}\label{sec:maybesoundness}

\begin{theorem}[Progress]\label{thm:progress}
 If \(\ \cx[\varphi]{[\varnothing\mid\Sigma]}  \ts t : \ty{Q}\) and \WF{\Sigma},
 then either \(t\) is a value, or for any store \(\sigma\) where
 \(\cx[\varphi]{[\varnothing \mid \Sigma]} \vdash \sigma\), there exists a term
 \(t'\) and a store \(\sigma'\) such that \(t \mid \sigma \to t' \mid \sigma'\).
\end{theorem}
\begin{proof}
By induction over the derivation \(\cx[\varphi]{[\varnothing\mid\Sigma]}  \ts t : \ty{Q}\).
\end{proof}

\noindent Similar to \cite{DBLP:journals/pacmpl/WeiBJBR24}, reduction preserves
types up to qualifier growth:

\begin{theorem}[Preservation]\label{thm:soundness}
  If \(\ \cx[\varphi]{[\varnothing\mid\Sigma]}  \ts t : \ty[q]{T}\),
  and \(\cx[\varphi]{[\varnothing \mid \Sigma]} \vdash \sigma\),
  and \(t \mid \sigma \to t' \mid \sigma'\),
  and \WF{\Sigma},
  then there exists \(\Sigma' \supseteq \Sigma\), \(\varphi' \supseteq \varphi \cup p\), and \(p \subq\DOM(\Sigma'\setminus\Sigma)\)
  such that \(\cx[\varphi']{[\varnothing \mid \Sigma']} \vdash \sigma'\)
  and \(\cx[\varphi']{[\varnothing\mid\Sigma']} \ts t' : \ty[{q[p/\qfresh]}]{T}\).
\end{theorem}
\begin{proof}
  We proceed by induction on the derivation
  \(\cx[\varphi]{[\varnothing\mid\Sigma]}  \ts t : \ty[q]{T}\).

  For \Cref{typing:t-app}, we distinguish two cases:

  1. If the argument qualifier is an empty or singleton qualifier, we apply the
  substitution lemma directly using the typing derivation from the hypothesis.

  2. If the argument qualifier is larger than a singleton qualifier, we first
  derive the non-occurrence assumption for the argument qualifier in the
  function return type. This ensures that the return qualifier remains preserved
  after stepping. We then construct a separate typing derivation for \(t\) with
  a qualifier \(w\) that is more precise than \(p\). In particular, if \(t\) is
  a location \(\ell\), then \(r = \ell\). Using this refined typing derivation,
  we apply the substitution lemma (\Cref{lem:subst_term}), thereby completing
  the case.  \end{proof}

\begin{corollary}[Preservation of Separation]\label{coro:preservation_separation}
Sequential reduction of two terms with disjoint qualifiers preserve types and
  disjointness:
\infrule{%
  \begin{array}{l@{\qquad}l@{\qquad}ll}
   \cx[\DOM(\Sigma)]{[\varnothing \mid \Sigma]} \ts t_1 : \ty[q_1]{T_1} &  t_1 \mid \sigma\phantom{'} \to t_1' \mid \sigma' & \varnothing \mid \Sigma \ts \sigma & \WF{\Sigma}\\[1ex]
\cx[\DOM(\Sigma)]{[\varnothing \mid \Sigma]} \ts t_2:\ty[q_2]{T_2} & t_2 \mid \sigma' \to t_2' \mid \sigma'' & q_1 \overlap q_2 \subq \{\vardiamondsuit\} &
  \end{array}
}{
  \begin{array}{ll@{\qquad}l@{\qquad}l}
\exists p_1\;p_2\;\Sigma'\;\Sigma''. & \cx[\DOM(\Sigma')\phantom{'}]{[\varnothing \mid \Sigma'\phantom{'}]} \ts t_1' : \ty[p_1]{T_1} & \Sigma'' \supseteq \Sigma' \supseteq \Sigma \\[1ex]
& \cx[\DOM(\Sigma'')]{[\varnothing \mid \Sigma'']} \ts t_2' : \ty[p_2]{T_2} & p_1 \overlap p_2 \subq \{\vardiamondsuit\}
  \end{array}
}
\end{corollary}
\begin{proof}
  By sequential application of Preservation (\Cref{thm:soundness}) and the fact that a reduction step
  increases the assigned qualifier by at most a fresh new location, thus preserving disjointness.
\end{proof}

\begin{corollary}[Progress and Preservation in Parallel Reductions]\label{coro:par_reduction}
  Non-value expressions with disjoint observability filters can be evaluated in parallel
  on non-overlapping parts of the store
  ($\sigma_{\restriction_{\varphi}}$ restricts the domain of $\sigma$ to
  locations in $\varphi$), and the resulting qualifiers remain separate:
  \vspace{-6pt}
\infrule{%
  \begin{array}{l@{\qquad}l@{\qquad}l@{\qquad}l}
   \cx[\varphi_1]{[\varnothing \mid \Sigma]} \ts t_1 : \ty[q_1]{T_1}
    & \cx[\varphi_1]{[\varnothing \mid \Sigma]} \vdash \sigma
    & t_1, t_2\text{ non-value}
    & \WF{\Sigma} \\[1ex]
   \cx[\varphi_2]{[\varnothing \mid \Sigma]} \ts t_2:\ty[q_2]{T_2}
    & \cx[\varphi_2]{[\varnothing \mid \Sigma]} \vdash \sigma
    & \varphi_1 \cap \varphi_2 \subq \varnothing & %
  \end{array}
}{
  \begin{array}{r@{\qquad}l@{\qquad}l@{\qquad}l}
    \exists \sigma_1'\;\sigma_2'\;\Sigma_1\;\Sigma_2\;p_1\;p_2\;\varphi_1'\;\varphi_2'. \\[1ex]
    t_1 \mid \sigma_{\restriction_{\varphi_1}} \to t_1' \mid \sigma_1'
      & \cx[\varphi_1']{[\varnothing \mid \Sigma_1]} \ts t_1' : \ty[p_1]{T_1}
      & \Sigma_1 \supseteq \Sigma \\[1ex]
    t_2 \mid \sigma_{\restriction_{\varphi_2}} \to t_2' \mid \sigma_2'
      & \cx[\varphi_2']{[\varnothing \mid \Sigma_2]} \ts t_2' : \ty[p_2]{T_2}
      & \Sigma_2 \supseteq \Sigma
      & p_1 \overlap p_2 \subq \{\vardiamondsuit\}
  \end{array}
}
\end{corollary}
\begin{proof}
  Since $\varphi_1$ and $\varphi_2$ are disjoint, $q_1$ and $q_2$ are also
  disjoint.  By Progress (\Cref{thm:progress}), $t_1$ and $t_2$ can be reduced
  to $t_1'$ and $t_2'$, respectively.  Then by Preservation
  (\Cref{thm:soundness}), the contractums are well-typed.
  With disjoint new locations picked for the two reductions, the resulting qualifiers $p_1$ and $p_2$ are also disjoint.
\end{proof}

Our shallow semantics preserves the same strict separation guarantees as the
deep model, without incurring the overhead of explicitly tracking transitively
reachable locations.

\paragraph{\textbf{Semantic Type Soundness}}
In addition to proving syntactic type soundness, we define unary logical relations for a variant of the $\maybelang{}$-calculus,
and prove semantic type soundness (the fundamental property).
The variant excludes type abstraction and does not include general subtyping rules, which have been modeled in prior work~\cite{bao_modeling_2023}.
However, it covers the subtyping for cyclic reference types.

With our cyclic reference types, programs in our system may not terminate.
Thus, we adapt \citet{bao_modeling_2023}'s semantic model by removing the termination property from the interpretation of function types and terms.
We then apply the techniques on step-indexed logical relations~\cite{ahmed2004semantics} and 
define worlds indexed by execution steps.
Interested readers can find our Rocq mechanization online. %
\section{Extension: Dual-Component References}\label{sec:dual}
\begin{figure}[t]\footnotesize
\begin{mdframed}

\judgement{Syntax}{}

\[\begin{array}{l@{\qquad}l@{\qquad}l}
T          & ::=  \cdots \mid \bot \mid \mty{}{\SRef{T}{q}{U}{p}}  \qquad  & \text{Types (with Bottom and Dual-Component Reference)} \\
\end{array}
\]

\judgement{Term Typing and Subtyping}{\BOX{\strut\G[\flt] \ts t : \ty{Q}}\ \BOX{\strut \G\ts\ty{Q} <: \ty{Q}}}\\

\begin{minipage}[t]{.44\textwidth}\small\vspace{0pt}
\typerule{t-sref-dual}{
    \G[\flt]\ts t : \ty[q]{T} \quad
    \QFresh\notin q
  }{
    \G[\flt]\ts \tref~t : \mty{\QFresh}{(\SRef{T}{q}{T}{q})}
  }
\end{minipage}
\begin{minipage}[t]{.01\textwidth}\end{minipage}
\begin{minipage}[t]{.46\textwidth}\small\vspace{0pt}
    \typerule{t-esc}{
      \G, x: \ty{R}\ts w <: q \quad
      \G, x: \ty{R}\ts s, p <: u, r \quad \G\ts p <: r \\ 
      r \subq \flt \quad \G[\flt]\ts t : \mty[x]{p}{(\SRef{T}{q}{U}{s})}
    }{
      \G[\flt]\ts t : \mty[x]{r}{(\SRef{T}{w}{U}{u, x})}
    }
\end{minipage}
\vspace{5pt}

\begin{minipage}[t]{.48\textwidth}\small\vspace{0pt}
    \typerule{t-sassgn-dual}{
      \G[\flt]\ts t_1 : \mty{p}{(\SRef{T}{q}{U}{s})} \\
      \G[\flt]\ts t_2 : \ty[q \ominus x]{T}
    }{
      \G[\flt]\ts t_1 \coloneqq t_2 : \ty[\qbot]{\TUnit}
    }
\end{minipage}
\begin{minipage}[t]{.01\textwidth}\end{minipage}
\begin{minipage}[t]{.48\textwidth}\small\vspace{0pt}
    \typerule{t-sassgn-v-dual}{
      \G[\flt]\ts y : \mty{p,y}{(\SRef{T}{q,x}{T}{s})} \\
      \G[\flt]\ts t_2 : \ty[q,y]{T}
    }{
      \G[\flt]\ts y \coloneqq t_2 : \ty[\qbot]{\TUnit}
    }
\end{minipage}

\vspace{5pt}

\begin{minipage}[t]{.6\textwidth}\small\vspace{0pt}
  \typerule{t-sassgn-l-dual}
  {
    \Sigma(\ell) = \mty[x]{q}{T} \quad
    \cx[\flt]{[\G\mid\Sigma]}\ts \ell : \mty{p,\ell}{\SRef{T}{u}{U}{s}}
    \\
    w \subq q \ominus x \quad
    \cx[\flt]{[\G\mid\Sigma]}\ts t_2 : \ty[{u[\{\ell,w\}/x]}]{T}
  }{
    \cx[\flt]{[\G\mid\Sigma]}\ts \ell \coloneqq t_2 : \ty[\qbot]{\TUnit}
  }
\end{minipage}
\begin{minipage}[t]{.05\textwidth}\end{minipage}
\begin{minipage}[t]{.3\textwidth}\small\vspace{0pt}
\typerule{t-sderef-dual}{
    \G[\flt]\ts t : \mty{p}{(\SRef{T}{q}{U}{s})} \\
    s\subq\flt,x \qquad x \notin \FV(U)
  }{
    \G[\flt]\ts !t : \ty[{s[p/x]}]{U}
  }
\end{minipage}

\vspace{5pt}

\begin{minipage}[t]{.2\textwidth}\small\vspace{0pt}
    \typerule{s-bot}{
      \\
      p\subq\DOM(\Gamma)
    }{
      \G\ts \ty[p]{\bot} <: \ty[p]{T}
    }
\end{minipage}
\begin{minipage}[t]{.05\textwidth}\end{minipage}
\begin{minipage}[t]{.68\textwidth}\small\vspace{0pt}
    \typerule{s-sref}{
      p\subq\DOM(\Gamma) \quad
      R = \mty{p}{(\SRef{T}{q}{U}{s})} \\
      \G, x: \ty{R}\ts w <: q \quad
      \G, x: \ty{R}\ts s <: u \quad
      \G\ts S <: T \quad
      \G\ts U <: V
    }{
      \G\ts \mty{p}{(\SRef{T}{q}{U}{s})} <: \mty{p}{(\SRef{S}{w}{V}{u})}
    }
\end{minipage}

\judgement{Notations}{}
\[\begin{array}{l@{\qquad}l@{\qquad}l@{\qquad}l}
  \mty{}{\TRef~[Q]}                     & := & \mty{}{\SRef{Q}{}{Q}{}}   & \text{Collapsed same qualified component} \\
  \mty{}{\TRef~[\rdcolor{\ty[\&p]{T}}]} & := & \mty{}{\SRf{T}{\qbot}{p}} & \text{Readonly references} \\
  \mty{}{\SRf{T}{q}{p}}                 & := & \mty{}{\SRef{T}{q}{T}{p}} & \text{Collapsed same type component} \\
\end{array}\]

\caption{Typing and subtyping rules for dual-component references in \maybelang.
The \wrcolor{write component} follows contravariance, while the \rdcolor{read
component} follows covariance.}\label{fig:dual-component-semantics}
\end{mdframed}
\vspace{-3ex}
\end{figure} 

This section extends reference typing in \maybelang{} with dual-component
references, introducing separate \wrcolor{write} and \rdcolor{read} components
to enable controlled imprecision and escaping.
\Cref{fig:dual-component-semantics} presents typing rules for constructing, escaping,
assigning, dereferencing, and subtyping such references.

\paragraph{Reference Introduction} \Cref{typing:t-sref-dual} introduces
dual-component references with separate \wrcolor{write} and \rdcolor{read}
components. Both components initially share the same type and qualifier,
ensuring maximum qualifier precision at allocation, but can later diverge via
subtyping (\Cref{typing:s-sref} and \Cref{typing:t-esc}).

\paragraph{Escaping} \Cref{typing:t-esc} introduces a mechanism for controlled 
imprecision under the shallow reference tracking model, where reference
qualifiers precisely track their own location without reaching their enclosed
referents. Escaping allows inner referent qualifiers to escape into the outer
reference qualifiers.  Operationally, the inner \rdcolor{read} qualifier can
partially or fully transfer its content to the outer qualifier. Under this rule,
although the \rdcolor{read} component is covariant, it can be weakened to a
\textit{smaller} qualifier after escaping. Such weakening is permitted as long
as the dropped portion is abstracted by the cyclic binder and transferred to the
outer qualifier. 

To ensure that the escaped component is properly tracked, the system adds a
cyclic binder to the escaped \rdcolor{read} component, enabling the outer
qualifier to be reintroduced upon dereferencing (See
\Cref{typing:t-sderef-dual}). Specifically, upon dereferencing an escaped
reference, the resulting qualifier will contain the union of the remaining inner
qualifier and the outer one, which includes all previously escaped resource.
Due to contravariance, the \wrcolor{write} component can only shrink but not
grow after escaping, similar to \Cref{typing:s-sref}. We do not add the cyclic
binder to the \wrcolor{write} component, so that escaped references will not
automatically allow cyclic assignment.

\paragraph{Assignment} \Cref{typing:t-sassgn-dual},
\Cref{typing:t-sassgn-v-dual}, and \Cref{typing:t-sassgn-l-dual} govern
assignment to dual-component references via the contravariant \wrcolor{write}
component. This ensures no value with a qualifier reaching more than that of its
initial referent qualifier is assigned. While subtyping (\Cref{typing:s-sref})
may grow the \rdcolor{read} component, it may only shrink the \wrcolor{write}
component.  Cyclic assignment (\Cref{typing:t-sassgn-v-dual}) is permitted only
for variables, mirroring \Cref{typing:t-sassgn-v}.  In
\Cref{typing:t-sassgn-l-dual}, we allow assignment to a location, similar to
\Cref{typing:t-sassgn-l}. However, due to the introduction of dual-component
references, where the \wrcolor{write} qualifier could have been shrunk due to
subtyping prior to the asignment. We use a store lookup to recover the original
referent qualifier, and permitting the assigned value to reach up to this
qualifier.

\paragraph{Bottom Type} \Cref{typing:s-bot} introduces the bottom type $\bot$ as
a subtype of all types, preserving qualifiers. It acts as a universal
placeholder for references with no write constraints.  Assigning $\bot$ to the
\wrcolor{write} component enforces immutability while preserving read access.
This also enables safe escape of nested references by abstracting away write
permissions without violating qualifier constraints.

\paragraph{Dereferencing} \Cref{typing:t-sderef-dual} handles dereferencing by 
using the \rdcolor{read} component, which is covariant and guaranteed to retain
its effective qualifier modulo any escaping. The cyclic binder in the read
qualifier is substituted with the reference's outer qualifier, as in
\Cref{typing:t-sderef}. 

\paragraph{Subtyping} \Cref{typing:s-sref} enforces variance rules for
dual-component references. The \wrcolor{write} component is contravariant,
preventing larger qualifiers from being assigned; the \rdcolor{read} component
is covariant, allowing safe expansion.

\paragraph{\textbf{Semantic Type Soundness}} 
We also prove semantic type soundness for the extension with dual-component references
by adapting \citet{bao_modeling_2023}'s semantic model and applying step-indexed logical relations~\cite{ahmed2004semantics}.
The details are elided for brevity, and interested readers can find our Rocq mechanization online. %
\section{Case Studies} \label{sec:casestudy}

Our system enables reasoning about safety and separation in the presence of
cyclic references, fine-grained parallelism, and controlled escapes. This
section explores three case studies demonstrating its versatility.

First, we show how a fixed-point combinator (\Cref{sec:casestudy-fixpoint}) can
be encoded as a well-typed program. Next, we examine fine-grained parallelism
(\Cref{sec:casestudy-shallow}), where shallow reference tracking enhances
concurrency. Finally, we introduce escaping, read-only references
(\Cref{sec:casestudy-escape}), allowing references to outlive their scope while
ensuring sound qualifier tracking.

\subsection{General Fixed-Point Combinator}
\label{sec:casestudy-fixpoint}
In this section, we present a case study demonstrating how cyclic reference
types enable the implementation of a general fixpoint operator, allowing
functions to define themselves recursively without explicit recursion.

Following \Citet{kiselyov_many_2020}, we implement a general fixpoint operator through the store as follows: 

\begin{lstlisting}
    def fix[T] (f: (g: (T -> T)$^\QFresh$ -> (T -> T)$^\texttt{g}$)) : (T -> T)$^\QFresh$ =
        val c = new Ref (x => x) // : $\lstcm{\mr[z]{c}{(T -> T)}{z}}$
        c := f((n:T) => (!c)(n)) // : (T -> T)$^\lstcm{\texttt{c}}$
        !c
\end{lstlisting}

The key idea is to use a reference to hold the function being defined
recursively, ensuring that it can be updated as needed. The fixpoint operator is
polymorphic over type @T@, making it applicable to recursive computations on
arbitrary data types. The function @f@ is a higher-order function passed as an
argument to the fixpoint operator, performing a ``one-step'' computation, using
its first argument @g@ (of type @T -> T@) as the function that it refines
recursively.

Unlike explicit recursive definitions, this approach enables recursion through
mutable, cyclic references, allowing functions to be defined incrementally. This
makes it particularly useful in settings where recursion needs to be established
dynamically, such as in languages without native support for recursion or when
defining mutually recursive functions.

We also implement an infinite loop and a factorial function as an instantiation
of the fixpoint operator, demonstrating more practical use cases.  
See \Cref{appendix:fixpoint} for details.

\subsection{Fine-Grained Parallelism}
\label{sec:casestudy-shallow}
\Citet{DBLP:journals/pacmpl/WeiBJBR24} treats a reference and all its  
transitively reachable objects as shared resources. This means that if a
reference @outer@ points to @inner@, then modifying either one is considered an
update to the same memory region. As a result, parallel updates to @inner@ and
@outer@ are disallowed, even if they are logically independent.  

Our system removes this limitation by introducing \textit{shallow reference  
tracking} via \rchty{}, which allows references pointing to shared store
locations to be treated as separate unless explicitly linked. This enables safe
parallel updates to different parts of a data structure.

To illustrate this, we extend the previous example by introducing an additional  
layer of references. Now, @inner1@ and @inner2@ are stored inside  
@outer2@, which is itself referenced by @outer1@. In \maybelang,  
@inner1@, @inner2@, @outer2@, and @outer1@ are  
considered separate, allowing concurrent updates to each:  

\begin{lstlisting}
    // Our system: inner1, inner2, outer2, and outer1 are separate.
    val inner1 = ... // : $\lstcm{\ty[\texttt{inner1}]{T}}$
    val inner2 = ... // : $\lstcm{\ty[\texttt{inner2}]{T}}$
    val outer2 = new Ref((inner1, inner2))  // : $\lstcm{\ty[\texttt{outer2}]{\TRef[(\ty[\texttt{inner1}]{T}, \ty[\texttt{inner2}]{T})]}}$
    val outer1 = new Ref(outer2) // : $\lstcm{\ty[\texttt{outer1}]{\TRef[\ty[\texttt{outer2}]{\TRef[(\ty[\texttt{inner1}]{T}, \ty[\texttt{inner2}]{T})]}]}}$
    def par(b1: (Unit => Unit)$^{\QFresh}$) (b2: (Unit => Unit)$^{\QFresh}$) = { ... // parallelize b1 and b2 }
    // Type checks: inner1, inner2, outer2, and outer1 are disjoint
    par { inner1 := ... } { outer1 := ... } // any permutation is allowed
\end{lstlisting}  

This approach enables \textit{fine-grained parallelism}, where multiple threads
or tasks can safely modify different parts of the structure  without
synchronization overhead. 
\subsection{Escaping, Read-Only References}
\label{sec:casestudy-escape}

In \maybelang, we introduce a principled approach that allows escaping
references while ensuring proper tracking of transitive reachability.

\paragraph{Ensuring Outer Observability via Forced Escaping}  

We demonstrate a case where a reference is forced to escape, ensuring that all
transitively reachable locations are tracked by the outer qualifier. This
recovers the deep semantics of \citet{DBLP:journals/pacmpl/WeiBJBR24} via
escaping.

\begin{lstlisting}
    // Function forcing a nested reference to escape
    def escapeNestedRef(r: $\mty[z]{\QFresh}{\SRef{\ensuremath{\bot}}{\qbot}{\mty[y]{}{\SRef{\ensuremath{\bot}}{\qbot}{\TRef[Int]}{y}}}{z}}$) = r // Return escaped reference
    val inner = new Ref(10)  // : $\lstcm{\ty[inner]{\TRef[Int]}}$
    val mid = new Ref(inner) // : $\lstcm{\ty[mid]{\TRef[\ty[inner]{\TRef[Int]}]}}$
    val ref = new Ref(mid)   // : $\lstcm{\ty[ref]{\TRef[\ty[mid]{\TRef[\ty[inner]{\TRef[Int]}]}]}}$
    escapeNestedRef(ref) 
    // : $\lstcm{\ty[ref]{\TRef[\ty[mid]{\TRef[\ty[inner]{\TRef[Int]}]}]} <: \mty[z]{inner,mid,ref}{\SRef{\ensuremath{\bot}}{\qbot}{\mty[y]{}{\SRef{\ensuremath{\bot}}{\qbot}{\TRef[Int]}{y}}}{z}}}$
    // Observes all reachable locations while allowing controlled escape
\end{lstlisting}

\paragraph{Strict Read-Only References via Reference Upcasting}  

Comparing to the example presented in \Cref{sec:motivation-escape}, a strict
form of read-only reference can be enforced by setting the write component to
the bottom type \(\bot\). This ensures that the reference is immutable, even
for untracked types:

\begin{lstlisting}
    def useImmutableRef(r: $\mty[z]{\QFresh}{\SRef{\ensuremath{\bot}}{\qbot}{\TRef[Int]}{\qbot}}$) =
        !r        // Allowed
    //  !!r := 42 // $\lsterrcm{\text{Error! Reference is fully immutable}}$
    val ref = new Ref(0) 
    useImmutableRef(ref) // Okay, upcasting to a fully immutable reference
    // : $\lstcm{\ty[r]{\TRef[Int]} <: \mty[z]{r}{\SRef{\ensuremath{\bot}}{z}{\TRef[Int]}{z}}}$
\end{lstlisting}

\noindent With this approach, the reference @r@ is treated as fully immutable  
(even when modified with an untracked value, because \(\bot\) is an uninhabited
type) at the abstraction boundary, while still allowing modifications in its
original context.

\section{Limitations and Future Work}
\label{sec:limitation}

This section outlines two categories of limitations. First, we discuss an intrinsic 
limitation of our approach in supporting expressive cyclic references 
(\Cref{sec:limitation-of}), namely its inability to form cycles across multiple 
``hops''. Second, we identify new restrictions introduced to the base system 
(\Cref{sec:limitation-new}), focusing on constraints in deep substitution for 
function applications. While necessary for soundness, these constraints reduce 
expressiveness in some cases.

\subsection{Limitations of Our Solution}\label{sec:limitation-of}

One limitation of our solution is that it cannot encode deeply nested cyclic
references containing multiple ``hops'' through the store, such as
\(\mr[z]{}{\mr[x]{}{T}{z}}{}\). Here, the inner referent's qualifier refers to
the qualifier for its outer parent $z$, rather than that of its immediate parent
$x$. This is not possible due to our introduction rule, which requires
cyclic-references to be tied to their immediate parent.

Supporting such multi-hop cycles poses challenges for well-formedness, 
qualifier tracking, and assignment semantics. References must remain 
well-scoped, qualifiers must propagate across levels, and assignments must 
preserve soundness.

One possible extension is \emph{simultaneous reference allocation}, allowing 
mutually dependent references to be introduced together. For example, we could
introduce a new introduction form that takes an arity argument specifying the
level of nesting, allowing multiple references to be allocated simultaneously.
Another possible approach is \emph{region-based allocation}, grouping references
under a shared qualifier, thus allowing multi-hop cycles.
Both would enable more expressive cyclic patterns while preserving soundness --
a direction we leave for future work.

\subsection{Limitations Introduced by Our Solution}\label{sec:limitation-new}

A key limitation introduced by our approach is that deep substitution is no longer 
fully general in non-fresh applications (\ie, \Cref{typing:t-app}). 
In prior systems, arguments could be deeply substituted into function bodies 
without constraint. In our system, this is restricted unless the argument's 
qualifier is empty or a singleton.

Specifically, in non-fresh applications, the argument must not appear in the 
return type. This mirrors the restriction already present in fresh applications 
(\Cref{sec:typing-app}), and ensures sound qualifier tracking by preventing 
unintended aliasing.

This is not unique to our design: Capturing Types 
\citep{DBLP:journals/toplas/BoruchGruszeckiOLLB23} impose the same restriction 
implicitly by requiring Monadic Normal Form (MNF), where function arguments are 
variables and not arbitrary expressions.

Despite the restriction, the impact on expressiveness is minor. As in DOT 
\citep{DBLP:conf/oopsla/RompfA16}, precision is lost only in specific cases, 
and a simple workaround exists: one can introduce an intermediate variable 
to recover deep substitution when needed.

\section{Related Work}
\label{sec:related}

\paragraph{\textbf{Reachability Types}}

\citet{DBLP:journals/pacmpl/BaoWBJHR21} introduced RT that support only
first-order mutable references. \citet{DBLP:journals/pacmpl/WeiBJBR24} extends
their work to support polymorphism and higher-order stores.  We address the
three key limitations in  \citet{DBLP:journals/pacmpl/WeiBJBR24}'s system.  We
introduce cyclic reference types, enabling recursive constructs without built-in
mechanisms.  Our system extends non-cyclic references to generalize the
semantics introduced in \citet{DBLP:journals/pacmpl/WeiBJBR24}, providing
additional flexibility for cyclic structures. 
We also refine the reference introduction rule to increase precision and allow
flexible graph structures with nested references (see
\Cref{sec:motivation-shallow}).

Following Dependent Object Types (DOT)
\citep{DBLP:conf/oopsla/RompfA16,DBLP:conf/birthday/AminGORS16}, RT functions
track their captured variables via self-references, ensuring precise tracking of
shared resources. In principle, function self-references could be used within
the function body to encode non-terminating recursions (\eg, \texttt{def f() =
f()}), making recursion a built-in feature. Instead, we explicitly enable cycles
through store locations, allowing cyclic references to model recursion without
having recursion as a primitive mechanism.

\Citet{bao_modeling_2023} formalize RT using logical relations to prove key
properties, including termination even in the presence of higher-order store,
which is a key premise for our work to build on. Graph IR
\citep{DBLP:journals/pacmpl/BracevacWJAJBR23} leverages RT to optimize impure higher-order programs by tracking fine-grained
dependencies with an effect system. \Citet{jia_escape_2024} address key
challenges in self-references by proposing an enhanced
notion of subtyping and developing a sound and decidable bidirectional typing
algorithm for RT. 

\paragraph{\textbf{Fixed-point Combinator}}

The origin of the fixed-point combinator can be traced to computability theory
\citep{rogers_theory_1967, DBLP:journals/toplas/Sangiorgi09}, which
characterizes the computational power of recursive functions.
\Citet{ProgrammingLanguagesandTheirDefinition:H.Bekic1936-1982/Bekic84}
formalizes recursion using fixed-point constructions, demonstrating that all
definable operations arise as fixed points of monotonic functions. 

Landin's Knot \citep{TheComputerJournal/Landin64} encodes mutual recursion
through self-application using mutable reference cells, a concept that later
influenced cyclic references in this work.
\Citet{Higher-OrderSymbComput/Goldberg05} generalizes Curry's fixed-point
combinator to handle variadic mutually-recursive functions.
\Citet{kiselyov_many_2020} delves into the diverse manifestations and
applications of fixed-point combinators in computation. Their approach is
particularly relevant to this work, as it provides a structured way to express
recursion through a store using cyclic references.

Our approach to cyclic references also draws inspiration from Landin's Knot, and
our case studies further explore the deep connection between recursion and
fixed-point constructs. While \citet{koronkevich_one_2022} restricts
non-termination in higher-order reference systems via environment
quantification, our work deliberately enables controlled non-termination,
leveraging cyclic references for recursion modeling.

\paragraph{\textbf{Separation}}

Separation Logic \citep{DBLP:conf/lics/Reynolds02} provides a formal framework
for local, spatial reasoning about mutable heap structures, utilizing a
separating conjunction operator to enforce memory disjointness.  RT
\citep{DBLP:journals/pacmpl/BaoWBJHR21,DBLP:journals/pacmpl/WeiBJBR24} 
incorporates principles of Separation Logic, particularly in its application
type rule (\Cref{typing:t-app-fresh}), where the overlap operator ensures
disjointness between function and argument qualifiers, maintaining separation
guarantees in type reasoning.  In this work, we reuse
\citep{DBLP:journals/pacmpl/WeiBJBR24}'s notion of separation expressed by the
overlap operator (see \Cref{fig:maybe-qtrans}) in \Cref{typing:t-app-fresh}.
Unlike previous approaches that rely on universally quantified saturated upper
bounds, we use an algorithmic transitive closure method to compute qualifier
overlap, yielding greater precision in qualifier reasoning (see
\Cref{sec:formal-mechanization}).

Ensuring static separation is crucial for safe concurrent programming.  Capture
Separation Calculus (CSC) \citep{xu_degrees_2023,Proc.ACMProgram.Lang./XuA24}
extends Capture Calculus (CC) \citep{DBLP:journals/toplas/BoruchGruszeckiOLLB23}
to enforce static separation and data race freedom, ensuring non-interference of
concurrently executing threads. \Citet{Proc.ACMProgram.Lang./DeVilhena21} extend
separation logic with built-in effect handlers to reason about cooperative
concurrency and stateful computations, guaranteeing safe access to
heap-allocated mutable state.

\paragraph{\textbf{Recursive Types and Existential Types}}

Alias Types \citep{Program.Lang.Syst./SmithD00} extends linear types with
aliasing to enable efficient and safe memory management, leveraging existential
types to abstract specific memory locations. \Citet{DBLP:conf/tic/WalkerM00}
employ existential types to capture memory shape information, particularly
focusing on the typing of recursive data structures through parameterized
recursive types that represent circular references.
\Citet{Fundam.Informaticae/AhmedM07} further utilize existential types to
abstract memory locations, whereas
\citet{Proc.ACMSIGPLAN2002Conf.Program.Lang.Des.Implement./GrossmanGTMY02a}
employ them to implement closures, where function pointers are paired with
existentially quantified environments to enforce region bounds and ensure region
liveness.

In contrast, our approach introduces a ``quasi-existential'' qualifier with a
semantics akin to that of recursive self types in DOT
\citep{DBLP:conf/birthday/AminGORS16,DBLP:conf/oopsla/RompfA16}, where the
identity of the parent reference is retained. We deliberately avoid existential
types in the context of cyclic references, as they fail to preserve the identity
of the cyclic reference, leading to a loss of precision. For example,
dereferencing (see \Cref{sec:typing-deref}) the same cyclic reference modeled by
existential types multiple times would produce values that appear separate, even
though they originate from the same reference.

\paragraph{\textbf{Regions, Ownership, and Substructural Type Systems}}

\Citet{DBLP:conf/pldi/MilanoTM22} introduce a system to guarantee separation and
safety in concurrent programs, combining linear typing, and region-based memory
management. However, while their work focuses on thread-level separation via
region types, RT address higher-order functions, providing fine-grained alias
tracking at the level of individual resources. 
\Citet{InformationandComputation/Tofte97} use regions to enforce sound life-time
management, proving semantic equivalence between source and target semantics,
ensuring the absence of use-after-free error. 
While their primary goal is a sound region inference algorithm for safe
deallocation, our work prioritizes soundness and separation, particularly in the
presence of cyclic references and recursive constructs. However, region-based
guarantees such as deallocation safety can still be achieved within RT by
layering an effect system \citep{DBLP:journals/pacmpl/BaoWBJHR21}.

\Citet{Proc.24thACMInt.WorkshopForm.Tech.Java-Programs/NobleJ23} propose local
ownership in Rust, allowing multiple mutable aliases within thread-local scopes
to support cyclic data. While their design loosens Rust's strict aliasing
discipline, it relies on lexical scoping and does not statically track heap
separation, which RT handles explicitly through qualifiers tracking.

\Citet{ECOOP2010-Object-OrientedProgram./Haller10} propose a capability-based
system that enforces at-most-once consumption of unique references while
allowing flexible borrowing. Linear Haskell
\citep{DBLP:journals/pacmpl/BernardyBNJS18} extends Haskell's type system with
linear types, providing fine-grained control over resource usage.
\Citet{Proc.ACMProgram.Lang./ArvidssonESSJM23b} present Reggio, a
capability-based region system enforcing isolation via a
single-window-of-mutability. While effective for concurrency, its stack-like
mutability discipline limits flexible aliasing across regions. RT, in contrast,
supports reasoning about aliasing in the presence of pervasive higher-order
functions and shared mutable state through heap reachability.

\paragraph{\textbf{Capabilities and Path Dependent Types, and Qualified Types}}

Dependent Object Types (DOT) \citep{DBLP:conf/oopsla/RompfA16} is formal model
of a subset of the Scala type system with proven soundness guarantee. One
analogous feature of DOT with this work is that it uses term-level
\textit{recursive self types}. DOT imposes restriction on dependent application,
where arguments are required to be of variable form, similar to
\Cref{typing:t-sassgn-v} in this work (see
\Cref{sec:motivation-cyclic,sec:formal-static}). However, recursive self types
in DOT are used to access member types, while RT use the self-referential
variable in cyclic reference for qualifier tracking.
\Citet{rapoportMutableWadlerFestDOT2017} extend DOT with mutable reference
cells, proving the soundness of their system. However, their types remain
unqualified, making it unsuitable for reachability tracking.
\Citet{LIPIcsVol.166ECOOP2020/Dort20} extend Dependent Object Types (DOT) with
reference mutability system to enable fine-grained access control on references
and provide immutability guarantees. \Citet{dort_pure_2024} focuses on side
effect free (SEF) function types, ensuring function purity statically.

Capturing Types (CT) \citep{DBLP:journals/toplas/BoruchGruszeckiOLLB23} is a
type system implemented in Scala, designed to track captured capabilities. While
CT makes use of boxing/unboxing operations, inspired by contextual modal type
theory (CMTT) \citep{DBLP:journals/tocl/NanevskiPP08}, RT's use of the \QFresh\
marker eliminates the need of boxing and unboxing, allowing for finer-grained
types that precisely model functions returning fresh values. 

Qualified types have been widely adopted to perform safety analysis such as
const qualifier inference in system-level languages
\citep{ACMTrans.Program.Lang.Syst./FosterRJ06}, reference immutability in Java
\citep{SIGPLANNot./HuangAW12}, and polymorphic type systems supporting
higher-order functions \citep{Proc.ACMProgram.Lang./Lee23}.

\section{Conclusion}%
\label{sec:conclusion}

In conclusion, this work presents a variant of the reachability types system
that extends its reference typing to allow cyclic references, precise and
shallow referent qualifier tracking, as well as escaping of reference qualifier
that enable expressive programming patterns such as fixed-point combinators,
safe parallelization, and read only references. 
\appendix

\section{Syntax for \maybelang} \label{appendix:syntax}

Based on \citet{DBLP:journals/pacmpl/WeiBJBR24}, \maybelang extends the
simply-typed \(\lambda\)-calculus with mutable cyclic references, qualified
types, and explicit type abstraction. The syntax uses metavariables \(x, y, z\)
for general term variables, \(f, g, h\) for function names or self-references,
and \(X\) for type variables.

Types \(T\) include the unit type \(\TUnit\), the top type \(\TTop\), type
variables \(X\), and qualified function types of the form \(f(x: Q) \to R\),
which bind both the function name \(f\) and the parameter \(x\) in the result
type \(R\). The function type may depend on both \(f\) and \(x\) via their
occurrence in qualifiers.  Types also include reference types
\(\mr{}{Q}{}\), and polymorphic types of the form \(\forall f(\ty[x]{X} <:
Q). Q\), which quantify over type variables \(X\) in a context that binds the
self-reference \(f\) and parameter \(x\).  Base types \(B\) are restricted to
\(\TUnit\), though this could be extended to include integers or booleans as
needed.

Terms \(t\) include constants \(c\), variables \(x\), recursive functions
\(\lambda f(x).t\), applications \(t~t\), reference allocations \(\tref~t\),
dereference operations \(!~t\), and assignments \(t \coloneqq t\).  The calculus
also supports type abstraction \(\TLam{X}{x}{T}{q}{t}\), which abstracts over a
type variable \(X\) with bound \(T\), and type application \(\TApp{t}{Q}{}\),
which instantiates a polymorphic term with a qualified type \(Q\).  Recursive
functions bind both the self-reference \(f\) and the argument \(x\), allowing
the body \(t\) to refer to both.

Reachability qualifiers \(p, q, r, w\) are finite sets of term and function
variables, and may include the freshness marker \(\QFresh\), which enforces
alias-tracking guarantees.  These qualifiers annotate types to control aliasing
and track reachability information, as in \(\ty[q]{T}\).  We refer to such
annotated types as qualified types \(Q, R\), while unqualified types are written
as \(T, U, V\).  In practice, we elide braces and write qualifiers as
comma-separated lists when convenient.

Observations \(\flt\) are finite sets of variables and are part of the term
typing judgment.  They specify the subset of variables in the typing environment
\(\Gamma\) that are considered observable.  The typing environment itself maps
term variables to qualified types \(x : Q\), and type variables to upper bounds
\(\ty[x]{X} <: Q\), maintaining a distinction between variable and type
assumptions.

\section{Unabridged figures corresponding to \Cref{sec:formal} in the main paper}

\subsection{Dynamic Semantics for \maybelang} \label{appendix:semantics}

Similar to \citet{DBLP:journals/pacmpl/WeiBJBR24}, we define the dynamic
semantics for \maybelang in a small-step operational semantics style. The
semantics is shown in \Cref{fig:maybe-reduction}.

A store and a store typing track the values, types, and qualifiers associated
with each location.  References reduce to fresh locations with their types and
qualifiers recorded in the store typing. Assignment statements evaluate to a
unit value while updating the store.  Function applications reduce to the
function body via $\beta$-reduction, replacing occurrences of the function
argument with a concrete value.  

\begin{figure}\small
\begin{mdframed}
\judgement{Reduction Contexts, Stores}{}
\[\begin{array}{l@{\ \ }c@{\ \ }l@{\qquad\qquad\ }l@{\ \ }c@{\ \ }l}
	{C} & ::= & \square \mid C\ t \mid v\ C \mid \tref~C \mid\ !{C} \mid {C} := {t} \mid {v} := {C} \mid C\ [Q] & \sigma & ::= & \varnothing \mid \sigma, \ell\mapsto v
\end{array}\]

\judgement{Reduction Rules}{\BOX{t \mid \sigma \to t \mid\sigma}}
\[\begin{array}{r@{\ \ }c@{\ \ }ll@{\qquad\qquad}r}
    \CX[gray]{C}{(\lambda f(x).t)\ v} \mid \sigma         & \to & \CX[gray]{C}{t[v/x, (\lambda f(x).t)/f]} \mid \sigma               &                           & \rulename{$\beta$}   \\
    \CX[gray]{C}{\tref~v} \mid \sigma                     & \to & \CX[gray]{C}{\ell} \mid (\sigma, \ell \mapsto v)                   & \ell \not\in \DOM(\sigma) & \rulename{ref}       \\
    \CX[gray]{C}{!\ell} \mid \sigma                       & \to & \CX[gray]{C}{\sigma(\ell)} \mid \sigma                             & \ell  \in \DOM(\sigma)    & \rulename{deref}     \\
    t
    \CX[gray]{C}{\ell := v} \mid \sigma                   & \to & \CX[gray]{C}{\tunit} \mid \sigma[\ell \mapsto v]                   & \ell \in \DOM(\sigma)     & \rulename{assign}    \\
    \CX[gray]{C}{(\Lambda f(\ty[x]{X}).t)\ Q} \mid \sigma & \to & \CX[gray]{C}t[Q/\ty[x]{X}, (\Lambda f(\ty[x]{X}).t)/f] \mid \sigma &                           & \rulename{$\beta_T$} \\
  \end{array}\]
\caption{Call-by-value reduction for \maybelang.}\label{fig:maybe-reduction}
\end{mdframed}
\end{figure}
 
\subsection{Transitive Closure Computation}

\begin{figure}\small
\begin{mdframed}
\judgement{Qualifier Cardinality}{\BOX{\cardinality{q}{\G}}}
\[\begin{array}{l@{\ \,}c@{\ \,}l@{\qquad\qquad\qquad\ \ }l}
	\cardinality{q}{\qbot}       & := & 0                       &                    \\ [1.1ex]
	\cardinality{q}{(\G, x : Q)} & := & 1 + \cardinality{q}{\G} & \text{if } x \in q \\ [1.1ex]
	\cardinality{q}{(\G, x : Q)} & := & \cardinality{q}{\G}     & \text{otherwise}   \\ [1.1ex]
\end{array}\]
\caption{Cardinality of Qualifiers}\label{fig:maybe-cardinality}
\end{mdframed}
\end{figure} 
We define the concept of cardinality on qualifiers to capture the number of
variables that is contained in a qualifier (see \Cref{fig:maybe-cardinality}).
Using cardinality, we establish a mapping between cardinality and qualifier
saturation. 

\begin{lemma}[Cardinality Monotonicity]\label{lem:cardinality-subq}
If $q_1 \subq q_2$,  
then $\cardinality{q_1}{\G} \leq \cardinality{q_2}{\G}$.
\end{lemma}
\begin{proof}
  By induction on the typing environment, with case analysis in the inductive
  case on whether the qualifier contains the top element.
\end{proof}

\begin{lemma}[Cardinality Max]\label{lem:cardinality-max}
$\cardinality{q}{\G} \leq \norm{\G}$.
\end{lemma}
\begin{proof}
  By induction on the typing environment, with case analysis in the inductive
  case on whether the qualifier contains the top element.
\end{proof}

\begin{lemma}[Zero Cardinality Saturated]\label{lem:cardinality-zero-sat}
If $\cardinality{q}{\G} = 0$,
Then $\sat{\G}{q}$.
\end{lemma}
\begin{proof}
  By induction on $\G$.
\end{proof}

\begin{lemma}[Cardinality and Sub-qualifier Preservation]\label{lem:cardinality-qor-trans}
If $\cardinality{q_1}{\G} = \cardinality{q_1 \qlub q_2}{\G}$,  
then ${\color{gray}\G\vdash}\, \qtrans[1]{q_1} \qlub\ \qtrans[1]{q_2} = \qtrans[1]{q_1} \qlub\ q_2$.
\end{lemma}
\begin{proof}
  By \Cref{lem:cardinality-subq}, we have $\cardinality{q_1}{\G} \leq
  \cardinality{q_1 \qlub q_2}{\G}$, By induction on the typing environment and
  with case analysis on whether the qualifiers contain the top element, in each
  case we either apply the induction hypothesis or derive a contradiction.
\end{proof}

\begin{lemma}[Transitive Lookup Monotone]\label{lem:trans-subq}
${\color{gray}\G\vdash}\, q \subq \qtrans[n]{q}$.
\end{lemma}
\begin{proof}
  By induction on $n$.
\end{proof}

\begin{lemma}[Cardinality Increment or Saturation]\label{lem:cardinality-inc-or-sat}
If ${\color{gray}\G\vdash}\, p = \qtrans[n]{q}$,
then $\cardinality{q}{\G} + n \leq \cardinality{p}{\G}$
or $\sat{\G}{p}$.
\end{lemma}
\begin{proof}
  We first prove the case for \( n = 1 \) and then proceed by induction on \( n \).  
  For the base case, we use \Cref{lem:trans-subq} and \Cref{lem:cardinality-subq} to establish the left disjunct.  
  For the inductive step, we apply the induction hypothesis and then invoke the case for \( n = 1 \).
\end{proof}

\begin{lemma}[Full Transitive Lookup Total]\label{lem:trans-full}
If $n \geq \norm{\G}$,
then ${\color{gray}\G\vdash}\, \qtrans[n]{q} = \qtrans[\norm{\G}]{q}$.
\end{lemma}
\begin{proof}
  Applying \Cref{lem:cardinality-inc-or-sat} to \( q \) with \( n \), we consider two cases:
  \begin{enumerate}
    \item Cardinality is increasing:  
      This contradicts \Cref{lem:cardinality-max}.
    \item \( q \) is saturated:  
      We analyze the cardinality value:
      \begin{itemize}
        \item If it is zero, we apply \Cref{lem:cardinality-zero-sat}.
        \item If it is nonzero, we apply \Cref{lem:cardinality-max} to derive a contradiction.
      \end{itemize}
  \end{enumerate}
\end{proof}
\section{Typing Rules for Natural Number and Boolean Extension} \label{appendix:nat}

This section presents the typing rules for the extension of the system with
natural numbers and booleans. The rules are shown in \Cref{fig:maybe-nat}.

\begin{figure}[t]\small
\begin{mdframed}
\begin{minipage}[t]{1.0\textwidth}\small
  \judgement{Syntax}{\BOX{\natlang}}\vspace{-8pt}
  \[\begin{array}{l@{\qquad}l@{\qquad}l@{\qquad}l}
    n &::=& 0 \mid 1 \mid \cdots & \text{Numeral Constants} \\
    b &::=& \ttrue \mid \tfalse & \text{Boolean Constants} \\
    T &::=& \cdots \mid \TNat \mid \TBool & \text{Types} \\
    t &::=& \cdots \mid \tnat{n} \mid \tsucc{t} \mid \tpred{t} \mid \tmul{t}{t} \mid \tiszero{t} & \\
    & & \qquad \mid \tbool{b} \mid  \tif{t}{t}{t} & \text{Terms} \\
    \Gamma &::=& \dots & \text{Typing Environments}
  \end{array}\]
  
  \judgement{Term Typing}{\BOX{\strut\G[\flt] \ts t : \ty{Q}}}\\[1ex]
  
\begin{minipage}[t]{0.48\linewidth}\small
    \typicallabel{t-nat}
    \typerule{t-nat}{}{
      \G[\varphi] \ts n : \ty[\qbot]{\TNat}
    }

    \vgap
    \typerule{t-succ}{
      \G[\varphi] \ts t : \ty[p]{\TNat}
    }{
      \G[\varphi] \ts \tsucc{t} : \ty[p]{\TNat}
    }
    \vgap
    \typerule{t-mul}{
      \G[\varphi] \ts t_1 : \ty[p]{\TNat} \quad
      \G[\varphi] \ts t_2 : \ty[q]{\TNat}
    }{
      \G[\varphi] \ts \tmul{t_1}{t_2} : \ty[p,q]{\TNat}
    }
\end{minipage}
\begin{minipage}[t]{0.48\linewidth}\small
    \typerule{t-iszero}{
      \G[\varphi] \ts t : \ty[p]{\TNat}
    }{
      \G[\varphi] \ts \tiszero{t} : \ty[\qbot]{\TBool}
    }

    \vgap
    \typerule{t-pred}{
      \G[\varphi] \ts t : \ty[p]{\TNat}
    }{
      \G[\varphi] \ts \tpred{t} : \ty[p]{\TNat}
    }
    
    \vgap
    \typerule{t-bool}{}{
      \G[\varphi] \ts b : \ty[\qbot]{\TBool}
    }

\end{minipage}
\begin{minipage}[t]{\linewidth}\small
    \typerule{t-if}{
      \G[\varphi] \ts t_1 : \ty[p]{\TBool} \qquad
      \G[\varphi] \ts t_2 : \ty[q]{T} \qquad
      \G[\varphi] \ts t_3 : \ty[r]{T}
    }{
      \G[\varphi] \ts \tif{t_1}{t_2}{t_3} : \ty[q,r]{T}
    }
\end{minipage}\\[1ex]

\end{minipage}\\[1ex]
  \judgement{Reduction Contexts, Values}{\ }\small
  \[
  \begin{array}{l@{\quad}l}
    {C} ::= \square \mid \tsucc{C} \mid \tpred{C} \mid \tmul{C}{t} \mid \tmul{t}{C} \mid \tiszero{C} \mid \tif{C}{t}{t} & \text{Reduction Contexts}\\
    {v} ::= \tnat{n} \mid \tbool{b} & \text{Values}
  \end{array}
  \]
 \judgement{Reduction Rules}{\BOX{t \mid \sigma \to t \mid \sigma}}\\[1ex]
  \[
  \begin{array}{r@{\ \ }c@{\ \ }ll@{\qquad\qquad}r}
      \CX{C}{\tsucc{\tnat{n}}} \mid \sigma & \to & \CX{C}{\tnat(n + 1)} \mid \sigma & & \rulename{succ} \\
      \CX{C}{\tmul{\tnat{n_1}}{\tnat{n_2}}} \mid \sigma & \to & \CX{C}{\tnat(n_1 \times n_2)} \mid \sigma & & \rulename{mul} \\
      \CX{C}{\tpred{\tnat{n}}} \mid \sigma & \to & \CX{C}{\tnat(\max(0, n-1))} \mid \sigma & & \rulename{pred} \\
      \CX{C}{\tiszero{\tnat{n}}} \mid \sigma & \to & \CX{C}{\tbool(n = 0)} \mid \sigma & & \rulename{iszero} \\
      \CX{C}{\tif{\tbool{\ttrue}}{t_1}{t_2}} \mid \sigma & \to & \CX{C}{t_1} \mid \sigma & & \rulename{if-true} \\
      \CX{C}{\tif{\tbool{\tfalse}}{t_1}{t_2}} \mid \sigma & \to & \CX{C}{t_2} \mid \sigma & & \rulename{if-false} \\
  \end{array}
  \]
\caption{Typing rules, reduction contexts, and reduction rules for natural
numbers and boolean expressions in \maybelang.}\label{fig:maybe-nat}
\end{mdframed}
\vspace{-3ex}
\end{figure}  %
\section{Details in the Fixed-Point combinator (\Cref{sec:casestudy})}\label{appendix:fixpoint}

We provide a detailed explanation of the fixed-point combinator, as presented in
\Cref{sec:casestudy} of the main paper. The body of the fixed-point combinator
performs the following steps:

\begin{enumerate}[topsep=0pt]

    \item \textit{Initializing the cyclic reference (\texttt{c}).} We create a
    reference \texttt{c} of type \texttt{T -> T}, initially holding the identity
    function \texttt{(x => x)} as a placeholder, ensuring type correctness and
    that the reference can be safely dereferenced before being updated.

    \item \textit{Defining the helper function (\texttt{f2}).} The function
    \texttt{f2} is a wrapper around \texttt{c}, retrieving its current value
    (\texttt{!c}) and applying it to the argument \texttt{n}. Since \texttt{c}
    will later be updated to hold the final recursive function, \texttt{f2}
    essentially acts as a proxy to dynamically retrieve and apply the most
    up-to-date version of the function.

    \item \textit{Updating \texttt{c} with the recursive function.} We apply
    \texttt{f} to \texttt{f2}, allowing \texttt{f} to construct the recursive
    function by invoking @f2@ as needed. The result of this computation
    (\texttt{f(f2)}) is then stored in \texttt{c}, ensuring @f2@ retrieves the
    updated definition at the next recursive step.

    \item \textit{Returning the fixpoint.} The fixpoint operator finally returns
    the dereferenced value of \texttt{c}, which now holds the fully-defined
    recursive function. Any subsequent calls to the function returned by
    \texttt{fix} will now correctly apply the recursively computed function.

\end{enumerate}

One possible instantiation of the fixpoint operator is an infinite loop. In this
case, the fixpoint operator is instantiated with type @Unit@, and the
``one-step'' computation is a function that unconditionally applies its first
argument to its second argument. 

\begin{lstlisting}
    def loop_fix[Unit](g: Unit -> Unit): Unit -> Unit = n => g(n)
    fix[Unit](loop_fix)(()) // infinite loop
\end{lstlisting}

To demonstrate more practical use cases than infinite loops, we extend the
system with integer types and basic arithmetic operations (see
\Cref{fig:maybe-nat}), enabling us to define a factorial function using the
fixpoint operator:

\begin{lstlisting}
    def fact_fix(f: Nat -> Nat): Nat -> Nat = 
        x => if (iszero x) then 1 else x * f (pred x)
    val n : Nat = ...
    fix[Nat](fact_fix)(n) // compute n!
\end{lstlisting}

Although in both case studies, we specialize the fixpoint operator with 
untracked types (\eg @Unit@ or @Nat@), the system supports reachability
polymorphism, allowing the fixpoint operator to be used with more complex types
, including closures. This extends its applicability beyond simple base types to
functions that capture references, enabling more advanced recursive
computations.

\begin{acks}                            %

This work was supported in part by NSF awards 2348334, Augusta faculty startup package, as
well as gifts from Meta, Google, Microsoft, and VMware.
\end{acks}

\section*{Data Availability Statement}
Rocq mechanizations can be found at \url{https://github.com/tiarkrompf/reachability}.

\bibliography{references}

\end{document}